\documentclass[english,reprint,superscriptaddress,aps]{revtex4-2}
\usepackage[T1]{fontenc}
\usepackage{textcomp}
\usepackage[latin9]{inputenc}
\usepackage{color}
\usepackage{babel}
\usepackage{verbatim}
\usepackage{mathrsfs}
\usepackage{amsmath}
\usepackage{stmaryrd}
\usepackage{graphicx}
\usepackage{tikz}
\usepackage[pdfusetitle,
 bookmarks=true,bookmarksnumbered=false,bookmarksopen=false,
 breaklinks=false,pdfborder={0 0 1},backref=false,colorlinks=true]
 {hyperref}

\makeatletter

\newcommand{\lyxmathsym}[1]{\ifmmode\begingroup\def\b@ld{bold}
  \text{\ifx\math@version\b@ld\bfseries\fi#1}\endgroup\else#1\fi}

\providecommand{\tabularnewline}{\\}

\makeatother

\begin{document}
\title{Quantum steering and discord in hyperon-antihyperon system in electron-positron annihilation}

\author{Sihao Wu}
\email{shwu@mail.ustc.edu.cn}
\affiliation{Department of Modern Physics and Anhui Center for fundamental sciences
in theoretical physics, University of Science and Technology of China,
Hefei 230026, China}

\author{Chen Qian}
\email{qianchen@baqis.ac.cn}
\affiliation{Beijing Academy of Quantum Information Sciences, Beijing 100193, China}

\author{Qun Wang}
\email{qunwang@ustc.edu.cn}
\affiliation{Department of Modern Physics and Anhui Center for fundamental sciences
in theoretical physics, University of Science and Technology of China,
Hefei 230026, China}
\affiliation{School of Mechanics and Physics, Anhui University of Science and Technology,
Huainan, Anhui 232001, China}

\author{Yang-Guang Yang}
\email{yyg@impcas.ac.cn}
\affiliation{Institute of Modern Physics, Chinese Academy of Sciences, Lanzhou
730000, China}

\begin{abstract}
Hyperon-antihyperon pairs produced in high-energy electron-positron
annihilation are promising systems for the study of quantum information
properties. In this work, we make an analysis of two types of quantum
correlations, the quantum steering and discord, in hyperon-antihyperon
systems produced in electron-positron annihilation based on the $X$-shaped spin density matrix. The behaviors of these
quantum correlations differ from those in elementary particle-antiparticle
systems such as the top quark and tau lepton due to the polarization
effect. The hierarchy of quantum correlations is examined and partially
confirmed in hyperon-antihyperon systems: $ \textrm{Bell Nonlocality} \subset
\textrm{Steering} \subset  \textrm{Entanglement} \subset \text{Discord}$. The loopholes and quantum decoherence effect are also discussed in our work.
\end{abstract}
\maketitle

\section{Introduction}

In recent years, testing quantum nonlocal correlations at colliders
has attracted a lot of interest in high-energy physics (HEP). Historically,
studies of fundamental problems in quantum mechanics were primarily
conducted in photonic and atomic systems \citep{Clauser:1974tg,Aspect:1981zz}.
However, with the progress in high-energy experiments, particularly
in particle detection technology and accumulation of huge amount of
experimental data, there has been a growing interest in the investigation
of quantum information properties at smaller length and higher energy
scales \citep{Barr:2024djo}.


Generally, the spin states of unstable particles produced in high
energy experiments which have weak decays can be extracted based on
the quantum state tomography \citep{Bernal:2023jba}. Recently, top
quarks as the heaviest of all known fundamental particles, serve as
an ideal testing ground for the study of quantum information in high
energy physics \citep{Afik:2020onf,Fabbrichesi:2021npl,Afik:2022kwm}.
The experimental evidence for the entanglement in $t\bar{t}$ has
been identified at Large Hadron Collider (LHC) as the entangled system
with the highest possible energy \citep{ATLAS:2023fsd}. Apart from
top quarks, investigation of other particle systems at colliders,
such as $\tau$ leptons \citep{Fabbrichesi:2022ovb,Ehataht:2023zzt,Han:2025ewp},
massive gauge bosons \citep{Barr:2021zcp,Barr:2022wyq,Aguilar-Saavedra:2022wam},
hyperons \citep{Tornqvist:1980af,Qian:2020ini,Wu:2024asu,Fabbrichesi:2024rec,Pei:2025yvr}, other quarks~\citep{Cheng:2025cuv} and vector mesons \citep{Fabbrichesi:2023idl}, have also been carried
out. Among these works, the quantum correlations in the hyperon-antihyperon
system in charmonium decays were proposed and can be tested at the
Beijing Spectrometer III (BESIII) \citep{Wu:2024asu,Wu:2024mtj,Fabbrichesi:2024rec,BESIII:2025vsr}. 


The Bell nonlocality (or violation of Bell inequality) and Bell entanglement
are a specific kind of the quantum correlation. As an asymmetric form
of the quantum correlation, the quantum steering or Einstein-Podolsky-Rosen
(EPR) steering was first introduced by Schr\"{o}dinger \citep{schrodinger:1935discussion}.
It refers to the ability that the measurement of one subsystem can
influence the state of another, even when the subsystems are spatially
separated. Steerability is regarded as a kind of quantum nonlocality
that is less restrictive than the Bell nonlocality \citep{Wiseman:2007prl,Uola:2020rmp}.

The quantum discord represents another form of the quantum correlation
\citep{Henderson:2001wrr,Ollivier:2001prl}. It captures the quantum nature of the correlation
that cannot be explained purely by classical physics, even when there
is no entanglement between the subsystems. It reflects more subtle,
yet fundamental, aspect of quantum mechanics, especially in the context
of quantum computation and quantum communication \citep{Datta:2008prl,Lanyon:2008prl,Dakic:2012nphys}.
Nowadays, the quantum steering and discord have been introduced into
particle physics such as in the neutrino oscillation \citep{Ming:2020nyc,Bittencourt:2022tcl}
and top quark systems \citep{Afik:2022dgh,Han:2024ugl}. 


In this work, we focus on the hyperon-antihyperon system produced
in $e^{+}e^{-}\to\gamma^{*}/\psi\to Y\bar{Y}$ and study the quantum
steering and discord in the $Y\bar{Y}$ system in a quantum information
perspective. Our investigation is mainly based on the two-qubit density
operator. We present the steerability of the system using the three-setting
measurement inequality \citep{Cavalcanti:2009pra,Costa:2016pra},
and give an analytical expression for the quantum discord from its
original definition \citep{Ollivier:2001prl}. In this way, we can
provide a full picture of quantum correlations including the Bell
nonlocality and entanglement given in our previous work \citep{Wu:2024asu}.
We examine the role of the hyperon\textquoteright s time-like
electromagnetic form factors (EMFFs) in shaping these quantum correlations, revealing how they lead to correlations distinct between the composed particles and the elementary ones. We also address two pressing experimental issues: the closure of loopholes in collider-based tests and the impact of quantum decoherence induced by the detectors.


The paper is organized as follows. In Sec. \ref{sec:Preliminaries},
we introduce the two-qubit density operator for hyperon-antihyperon
in $e^{+}e^{-}$ annihilation. We investigate the quantum steering
and discord in Sec. \ref{sec:Quantum-Steering} and \ref{sec:Quantum-Discord}
respectively. In Sec. \ref{sec:Hierarchy-of-quantum}, a hierarchy
of quantum correlations is presented. The role of hyperon's time-like
electromagnetic form factors is discussed in Sec.~\ref{sec:EMFFs}. The nonlocality loophole and quantum decoherence are addressed in Sec.~\ref{sec:loophole} and \ref{sec:decoherence}.
In Sec.~\ref{sec:Summary}, a summary of the main results and an outlook
are presented.


\section{Preliminaries \label{sec:Preliminaries}}

We start from the spin density operator for the hyperon and antihyperon
\citep{Faldt:2017kgy,Perotti:2018wxm} as a two-qubit system,
\begin{align}
\rho_{Y\bar{Y}}= & \frac{1}{4}\biggl(1\otimes1+\mathbf{B}^{+}\cdot\boldsymbol{\sigma}\otimes1+1\otimes\mathbf{B}^{-}\cdot\boldsymbol{\sigma}\nonumber \\
 & +\sum_{i,j}C_{ij}\sigma_{i}\otimes\sigma_{j}\biggr),\label{eq:two_qubit}
\end{align}
where $\boldsymbol{\sigma}=(\sigma_{1},\sigma_{2},\sigma_{3})$ are
Pauli matrices, $\mathbf{B}^{+}$ and $\mathbf{B}^{-}$ are the spin
polarization vectors for the hyperon ($Y$) and antihyperon ($\bar{Y}$)
respectively, and $C_{ij}$ is the 3$\times$3 spin correlation matrix
between them. In the analytical calculation with CP symmetry, the
spin polarization vectors of the hyperon and antihyperon have the
form
\begin{equation}
\mathbf{B}^{+}=\mathbf{B}^{-}=\left(0,\frac{\beta_{\psi}\sin\vartheta\cos\vartheta}{1+\alpha_{\psi}\cos^{2}\vartheta},0\right),
\end{equation}
and the spin correlation matrix is 
\begin{eqnarray}
C_{ij} & = & \frac{1}{1+\alpha_{\psi}\cos^{2}\vartheta}\nonumber \\
 &  & \times\begin{bmatrix}\sin^{2}\vartheta & 0 & \gamma_{\psi}\sin\vartheta\cos\vartheta\\
0 & -\alpha_{\psi}\sin^{2}\vartheta & 0\\
\gamma_{\psi}\sin\vartheta\cos\vartheta & 0 & \alpha_{\psi}+\cos^{2}\vartheta
\end{bmatrix},
\end{eqnarray}
where $\alpha_{\psi}\in[-1,1]$ is the decay parameter, $\Delta\Phi\in(-\pi,\pi]$
is the relative phase, $\vartheta$ is the scattering angle between
the electron beam and the outgoing hyperon, $\beta_{\psi}$ and $\gamma_{\psi}$
are defined as 
\begin{align}
\beta_{\psi}= & \sqrt{1-\alpha_{\psi}^{2}}\sin(\Delta\Phi),\nonumber \\
\gamma_{\psi}= & \sqrt{1-\alpha_{\psi}^{2}}\cos(\Delta\Phi).
\end{align}
The two parameters are measured at BESIII, and we listed some of the
results in Table \ref{tab:decay_parameters}. Here $\mathbf{B}^{\pm}$
and $C_{ij}$ are defined in hyperon's and antihyperon's rest frames.
After that, we choose three axes to be 
\begin{equation}
\hat{\mathbf{y}}=\frac{\hat{\mathbf{p}}_{Y}\times\hat{\mathbf{p}}_{e}}{|\hat{\mathbf{p}}_{Y}\times\hat{\mathbf{p}}_{e}|},\;\hat{\mathbf{z}}=\hat{\mathbf{p}}_{Y},\;\hat{\mathbf{x}}=\hat{\mathbf{y}}\times\hat{\mathbf{z}},
\end{equation}
which applies to both the hyperon and antihyperon.


From local unitary equivalence \citep{Wu:2024asu}, the quantum correlation
in a bipartite system keeps invariant under the local unitary transformation,
so we can freely convert the two-qubit density operator to a standard
$X$ state, 
\begin{align}
\rho_{Y\bar{Y}}^{X}= & \frac{1}{4}\biggl(1\otimes1+a\sigma_{z}\otimes1+1\otimes a\sigma_{z}\nonumber \\
 & +\sum_{i}t_{i}\sigma_{i}\otimes\sigma_{i}\biggr),\label{eq:X_state}
\end{align}
where 
\begin{align}
a & =\frac{\beta_{\psi}\sin\vartheta\cos\vartheta}{1+\alpha_{\psi}\cos^{2}\vartheta},\nonumber \\
t_{1,2} & =\frac{1+\alpha_{\psi}\pm\sqrt{(1+\alpha_{\psi}\cos2\vartheta)^{2}-(\beta_{\psi}\sin2\vartheta)^{2}}}{2(1+\alpha_{\psi}\cos^{2}\vartheta)},\nonumber \\
t_{3} & =\frac{-\alpha_{\psi}\sin^{2}\vartheta}{1+\alpha_{\psi}\cos^{2}\vartheta}.\label{eq:at123}
\end{align}
From Eqs.~(\ref{eq:two_qubit}) and (\ref{eq:X_state}), we can see
that the density matrices are locally unitary to each other, and the
$X$-state may dramatically simplify the analysis in following sections.
As shown in Eq.~(\ref{eq:at123}), the $X$-form density operator~(\ref{eq:X_state}) depends on two parameters $(\alpha_{\psi},\Delta\Phi)$
as a function of $\vartheta$. By explicitly express the Pauli matrices
in (\ref{eq:X_state}) following Ref. \citep{Wu:2024asu}, the $X$-state
$\rho_{Y\bar{Y}}^{X}$ can be put into the form

\begin{equation}
\rho_{Y\bar{Y}}^{X}=\frac{1}{4}\begin{bmatrix}1+2a+t_{3} & 0 & 0 & t_{1}-t_{2}\\
0 & 1-t_{3} & t_{1}+t_{2} & 0\\
0 & t_{1}+t_{2} & 1-t_{3} & 0\\
t_{1}-t_{2} & 0 & 0 & 1-2a+t_{3}
\end{bmatrix},\label{eq:X_form}
\end{equation}
where the name $X$-state comes from its resemblance to the letter
$X$.

\begin{table*}
\caption{\label{tab:decay_parameters}Some parameters in $e^{+}e^{-}\rightarrow J/\psi\rightarrow Y\bar{Y}$,
where $Y\bar{Y}$ is a pair of ground-state octet hyperons.}

\begin{ruledtabular}
\begin{tabular}{ccccc}
 & Branching Ratio ($\times10^{-4}$) & $\alpha_{\psi}$ & $\Delta\Phi/\mathrm{rad}$ & Ref\tabularnewline
\hline 
$J/\psi\to\Lambda\bar{\Lambda}$ & $19.43\pm0.03\pm0.33$ & $0.4748\pm0.0022\pm0.0031$ & $0.7521\pm0.0042\pm0.0066$ & \citep{BESIII:2018cnd,BESIII:2017kqw}\tabularnewline
$J/\psi\to\Sigma^{+}\bar{\Sigma}^{-}$ & $10.61\pm0.04\pm0.36$ & $-0.508\pm0.006\pm0.004$ & $-0.270\pm0.012\pm0.009$ & \citep{BES:2008hwe,BESIII:2020fqg}\tabularnewline
$J/\psi\to\Sigma^{0}\bar{\Sigma}^{0}$ & $11.64\pm0.04\pm0.23$ & $-0.4133\pm0.0035\pm0.0077$ & $-0.0828\pm0.00\pm0.0033$ & \citep{BESIII:2024nif}\tabularnewline
$J/\psi\to\Xi^{-}\bar{\Xi}^{+}$ & $10.40\pm0.06\pm0.74$ & $0.586\pm0.012\pm0.010$ & $1.213\pm0.046\pm0.016$ & \citep{BESIII:2021ypr,ParticleDataGroup:2022pth}\tabularnewline
$J/\psi\to\Xi^{0}\bar{\Xi}^{0}$ & $11.65\pm0.04\pm0.43$ & $0.514\pm0.006\pm0.0015$ & $1.168\pm0.019\pm0.018$ & \citep{BESIII:2016nix,BESIII:2023drj}\tabularnewline
\end{tabular}
\end{ruledtabular}

\end{table*}

\section{Quantum Steering \label{sec:Quantum-Steering}}

Quantum steering, also known as EPR steering, was first proposed in
1935 \citep{schrodinger:1935discussion}. It was reformulated in the
modern quantum information theory by \citep{Wiseman:2007prl} in 2007.
Similar to the local-hidden-variable (LHV) hypothesis which is introduced
in the Bell inequality, the local-hidden-state (LHS) model is put
forward for the steerability of quantum systems.


In the context of a two-qubit state $\rho_{AB}$ shared by Alice and
Bob, the quantum steering means that Alice can ``steer'' Bob\textquoteright s
state \textit{iff} (if and only if) the measurement outcomes of Alice
and Bob exhibit correlation that violates the LHS model \citep{Wiseman:2007prl}.
This violation suggests that the quantum system demonstrates a nonlocal
property that cannot be accounted for by any LHVs, thereby distinguishing
the quantum steering from classical correlations.


\subsection*{Detecting steerability via steering inequality}

Similar to the Bell nonlocality, the quantum steering involves the
violation of certain inequalities (known as steering inequalities)
that serve as its criterion. If these inequalities are violated, it
indicates that the system exhibits a genuine quantum behavior. Cavalcanti,
Jones, Wiseman, and Reid developed an inequality known as the CJWR
steering inequality to diagnose the steerability of a two-qubit state
with three-setting measurements on each side \citep{Cavalcanti:2009pra,Costa:2016pra},
\begin{equation}
F_{3}^{\mathrm{CJWR}}\equiv\frac{1}{\sqrt{3}}\left|\sum_{k=1}^{3}\mathrm{Tr}\left[\rho\left(A_{k}\otimes B_{k}\right)\right]\right|\leq1,\label{eq:CJWR}
\end{equation}
where $A_{k}=\mathbf{s}_{k}\cdot\boldsymbol{\sigma}$, $B_{k}=\mathbf{r}_{k}\cdot\boldsymbol{\sigma}$,
$\mathbf{s}_{k}$ and $\mathbf{r}_{k}$ $(k=1,2,3)$ are unit vectors,
and $\{\mathbf{r}_{1},\mathbf{r}_{2},\mathbf{r}_{3}\}$ are three
basis vectors in a Cartesan coordinate system (any two vectors are
orthogonal). 

\begin{figure}
\includegraphics[scale=0.49]{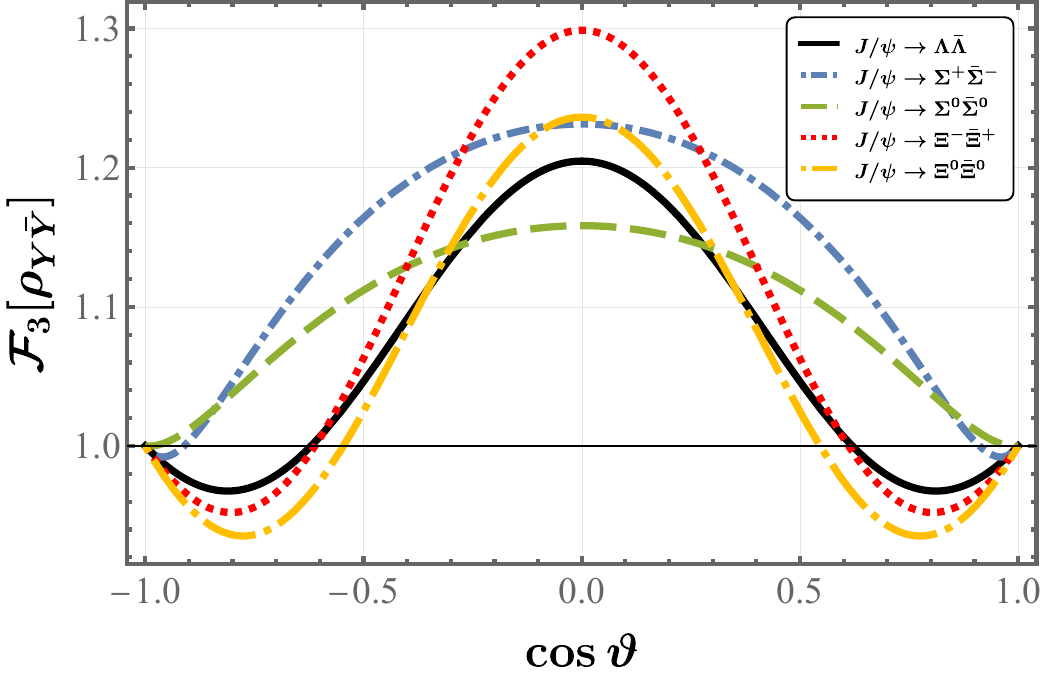}

\caption{\label{fig:F3}The quantity $\mathcal{F}_{3}[\rho_{Y\bar{Y}}]$ for
the three-setting measurement as functions of $\cos\vartheta$ ($\vartheta$
is the scattering angle) in $e^{+}e^{-}\to J/\psi\to Y\bar{Y}$ with
the black solid, blue dash-dotted, green dashed, red dotted and long
yellow dash-dotted lines representing $Y=\Lambda$, $\Sigma^{+}$,
$\Sigma^{0}$, $\Xi^{-}$ and $\Xi^{0}$ respectively. The black horizontal
line is the steering bound $\mathcal{F}_{3}=1$. The CJWR inequality
is violated iff $\mathcal{F}_{3}>1$. }
\end{figure}

By setting the unit vectors $\mathbf{r}_{k}$ as $||C\mathbf{r}_{1}||=||C\mathbf{r}_{2}||=||C\mathbf{r}_{3}||=\sqrt{\mathrm{Tr}(C^{\mathrm{T}}C)/3}$
and $\mathbf{s}_{k}=C\mathbf{r}_{k}/\sqrt{\mathrm{Tr}(C^{\mathrm{T}}C)/3}$,
$F_{3}^{\mathrm{CJWR}}$ in Eq. (\ref{eq:CJWR}) reaches its maximum
\citep{Du:2021pra}, where $C_{ij}$ is the $3\times3$ correlation
matrix in Eq. (\ref{eq:two_qubit}). Then the consequent maximal violation
reads
\begin{equation}
\mathcal{F}_{3}[\rho]\equiv\max_{\mathbf{s}_{k},\mathbf{r}_{k}}F_{3}^{\mathrm{CJWR}}=\sqrt{\mathrm{Tr}\left(C^{\mathrm{T}}C\right)}.
\end{equation}
By applying the inequality to the $X$-shape density operator for
$Y\bar{Y}$, we obtain the maximal violation of the CJWR steering
inequality
\begin{align}
 & \mathcal{F}_{3}\left[\rho_{Y\bar{Y}}\right]=\sqrt{t_{1}^{2}+t_{2}^{2}+t_{3}^{2}}\nonumber \\
 & =\sqrt{1+2\left(\frac{\alpha_{\psi}\sin^{2}\vartheta}{1+\alpha_{\psi}\cos^{2}\vartheta}\right)^{2}-2\left(\frac{\beta_{\psi}\sin\vartheta\cos\vartheta}{1+\alpha_{\psi}\cos^{2}\vartheta}\right)^{2}},\label{eq:F3}
\end{align}
which is shown in Fig. \ref{fig:F3} for a variety of hyperons. We
see that $\mathcal{F}_{3}[\rho_{Y\bar{Y}}]$ is a function of $\cos\vartheta$
($\vartheta$ is the scattering angle) and is symmetric with respect
to $\vartheta=90\lyxmathsym{\textdegree}$, with the black horizontal
line being the steering bound $\mathcal{F}_{3}=1$. 


There is a clear evidence that all hyperon-antihyperon systems in
$e^{+}e^{-}\to J/\psi\to Y\bar{Y}$ violate the CJWR steering inequality.
Furthermore, the violation has a peak in the transverse scattering
at $\vartheta=90^{\circ}$ with the maximal value $\mathcal{F}_{3}=\sqrt{1+2\alpha_{\psi}^{2}}$
which depends on the decay parameter $\alpha_{\psi}$ only. So we
say that the steerablity for hyperon-antihyperon systems can be quantified
through the violation of the CJWR inequality. The ranges of steerability
$(\vartheta^{*},\pi-\vartheta^{*})$ for hyperon-antihyperon systems
depend on both $\alpha_{\psi}$ and $\Delta\Phi$, where the critical
angle $\vartheta^{*}$ is given by 
\begin{equation}
\vartheta^{*}=\arctan\left|\sqrt{1-\alpha_{\psi}^{2}}\frac{\sin\Delta\Phi}{\alpha_{\psi}}\right|.
\end{equation}
The maximal violation for $\Sigma$ is weaker than that for $\Xi$
due to the relative small $\alpha_{\psi}$, while the range of steerability
for $\Sigma$ is broader due to the very small value of $\Delta\Phi$. 


It should be noted that the CJWR inequality serves only as a sufficient
condition for the quantum steering. Consequently, states that do not
violate the inequality are not necessarily unsteerable. Apart from
the CJWR inequality, there are other criteria for the quantum steering
as well \citep{Uola:2020rmp}. For instance, an alternative approach
to detect the quantum steering is presented in Appendix \ref{sec:steering_en}
through the entanglement \citep{Das:2019pra,Zhang:2021pra}. However,
most available criteria are sufficient but not necessary, limiting
their ability to detect steerable states.


So far, the only known necessary and sufficient criterion for the
quantum steering was introduced in Ref. \citep{Nguyen:2019prl} via
the definition of the\textit{ critical radius}. A rigorous proof is
given that a two-qubit state is steerable \textit{iff} the critical
radius is less than 1. This criterion has been applied to the top
quark system \citep{Afik:2022dgh}. However, a closed formula for
the critical radius exists only for the Bell diagonal state, limiting
its applicability to hyperon-antihyperon systems. 


\section{Quantum Discord \label{sec:Quantum-Discord}}

The quantum discord is another kind of the quantum property that quantifies
how much quantum correlation shared by a bipartite system. It needs
the introduction of the \textit{quantum mutual information} of a bipartite
system $\rho_{AB}$ defined as 
\begin{align}
\mathcal{I}(A:B) & \equiv S(\rho_{A})+S(\rho_{B})-S(\rho_{AB})\nonumber \\
 & =S(\rho_{A})-S(\rho_{A|B}),\label{eq:mutal}
\end{align}
where $S(\rho)\equiv-\mathrm{Tr}(\rho\log_{2}\rho)$ denotes the von
Neumann entropy, and $S(\rho_{A|B})\equiv S(\rho_{B})-S(\rho_{AB})$
is the conditional entropy.


To quantify the quantum discord, one can use a set of one-dimensional
projectors \citet{Ollivier:2001prl}. Let $\left\{ \Pi_{k},\sum_{k}\Pi_{k}=1\right\} $
denote a set of projectors that perform the projective measurement
on subsystem $B$. The post-measurement state of the subsystem $A$
conditional on the measurement outcome $k$ of $B$ reads
\begin{equation}
\rho_{A|\Pi_{k}}=\frac{1}{p_{k}}\mathrm{Tr}_{B}\left[(1\otimes\Pi_{k})\rho_{AB}(1\otimes\Pi_{k})\right],
\end{equation}
where the probability $p_{k}$ is given by $\mathrm{Tr}[(1\otimes\Pi_{k})\rho_{AB}(1\otimes\Pi_{k})]$.
The post-measurement states of $A$ form an ensemble $\left\{ p_{k},\rho_{A|\Pi_{k}}\right\} $.
Specifically, in a qubit bipartite system, $k=0,1$. Afterwards, the
mutual information conditional on the projective measurement performed
on the subsystem $B$ is defined as 
\begin{equation}
\mathcal{J}(A:B)\equiv S(\rho_{A})-\sum_{k}p_{k}S(\rho_{A|\Pi_{k}}),\label{eq:c_mutual}
\end{equation}
which is similar to Eq. (\ref{eq:mutal}). In order to get rid of
the measurement dependence, the classical mutual information of the
state $\rho_{AB}$ is given as maximizing the value of $\mathcal{J}(A:B)$
over all possible projectors. 


The quantum discord is defined as the difference between the quantum
and classical mutual information (\ref{eq:mutal}) and (\ref{eq:c_mutual}),
\begin{align}
\mathcal{D}[\rho_{AB}] & \equiv\mathcal{I}(A:B)-\max_{\{\Pi_{k}\}}\mathcal{J}(A:B)\nonumber \\
 & =S(\rho_{B})-S(\rho_{AB})+\min_{\{\Pi_{k}\}}\sum_{k}p_{k}S(\rho_{A|\Pi_{k}}).\label{eq:discord}
\end{align}
From its definition, one can see that the quantum discord is an entropy-like
quantity, which is non-negative and can never exceed one, i.e. $0\leq\mathcal{D}[\rho_{AB}]\leq1$.
Another feature of the quantum discord is that it is an asymmetric
quantity --- performing the measurement on $A$ rather than on $B$
returns, in general, a different value. However, in our work all hyperon-antihyperon
states at $e^{+}e^{-}\to Y\bar{Y}$ are symmetric due to $\mathcal{CP}$
symmetry (\ref{eq:X_state}), ensuring identical values of $\mathcal{D}[\rho_{Y\bar{Y}}]$
regardless of whether the measurement is performed on $Y$ or $\bar{Y}$. 


\subsection*{Discord for rank-$2$ $X$ states}

The obstacle to obtain the quantum discord lies in the sophisticated
maximization (or minimization) procedure outlined in Eq\@. (\ref{eq:discord}),
which must be done over all possible projective measurements on $B$.
The difficulty makes the exact expression of the quantum discord for
general two-qubit states still a blank. 


We know that the spin states of hyperon-antihyperon systems in $e^{+}e^{-}$
scattering are \textit{symmetric rank-$2$ $X$ states} with two non-zero eigenvalues~\citep{Wu:2024asu}:
\begin{equation}
\lambda_{1,2}=\frac{1}{2}\left(1\mp\frac{\alpha_{\psi}\sin^{2}\vartheta}{1+\alpha_{\psi}\cos^{2}\vartheta}\right).
\label{eq:eigenvalues}
\end{equation}
Fortunately, there is an exact and analytical formula for the quantum
discord of $X$ states \citep{Fanchini:2010pra,Shi:2011jpa,Jing:2016jpa,Zhu:2018qip}
\begin{eqnarray}
\mathcal{D}\left[\rho_{X}\right] & = & 1-\frac{1+a}{2}\log_{2}\frac{1+a}{2}-\frac{1-a}{2}\log_{2}\frac{1-a}{2}\nonumber \\
 &  & +\sum_{i=1}^{4}\lambda_{i}\log_{2}\lambda_{i}-\max_{\epsilon\in[0,1]}F(\epsilon),\label{eq:X-discord}
\end{eqnarray}
where $F(\epsilon)$ is defined as 
\begin{eqnarray}
F(\epsilon) & = & \frac{1+a\epsilon+H_{+}}{4}\log_{2}\frac{1+a\epsilon+H_{+}}{1+a\epsilon}\nonumber \\
 &  & +\frac{1+a\epsilon-H_{+}}{4}\log_{2}\frac{1+a\epsilon-H_{+}}{1+a\epsilon}\nonumber \\
 &  & +\frac{1-a\epsilon+H_{-}}{4}\log_{2}\frac{1-a\epsilon+H_{-}}{1-a\epsilon}\nonumber \\
 &  & +\frac{1-a\epsilon-H_{-}}{4}\log_{2}\frac{1-a\epsilon-H_{-}}{1-a\epsilon},
\end{eqnarray}
with $H_{\pm}=\sqrt{t^{2}(1-\epsilon^{2})+(a\pm t_{3}\epsilon)^{2}}$,
$t=\max\left\{ |t_{1}|,|t_{2}|\right\} $. Equation (\ref{eq:X-discord})
accounts for all two-qubit $X$ states, and according to Ref\@. \citep{Jing:2016jpa},
the quantum discord of any rank-$2$ mixed state of $X$-type are
always given by either $F(1)$ or $F(0)$. Simplifying Eq. (\ref{eq:X-discord})
and using the numerical method, we find the discord for $Y\bar{Y}$
in our study can always be obtained at $\epsilon=0$, so we have 
\begin{eqnarray}
\mathcal{D}[\rho_{Y\bar{Y}}] & = & h\left(\frac{1+a}{2}\right)-h\left(\frac{1+t_{3}}{2}\right)\nonumber \\
 &  & +h\left(\frac{1+\sqrt{t_{1}+t_{3}-t_{1}t_{3}}}{2}\right),
\end{eqnarray}
where 
\begin{equation}
h(x)\equiv-x\log_{2}x-(1-x)\log_{2}(1-x),
\end{equation}
is the Shannon binary entropy function, and $a$ and $t_{1,2,3}$
are functions of $\vartheta$ which depend on parameters $\alpha_{\psi}$
and $\Delta\Phi$ as shown in Eq. (\ref{eq:at123}). 


\begin{figure}
\includegraphics[scale=0.49]{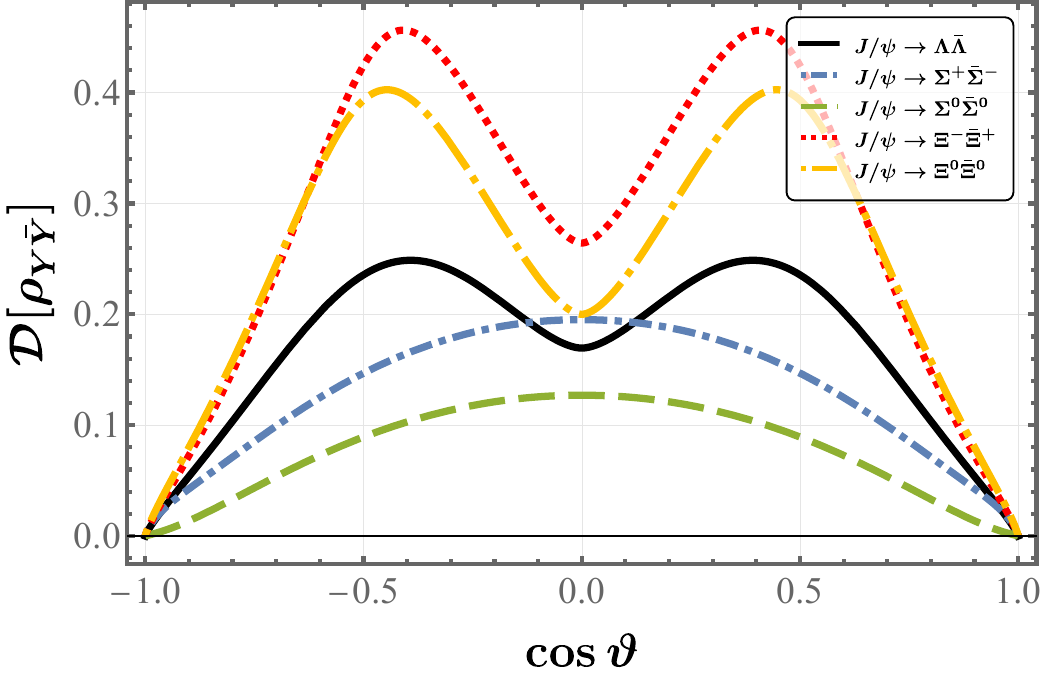}

\caption{\label{fig:discord}The quantum discord $\mathcal{D}\left[\rho_{Y\bar{Y}}\right]$
as functions of $\cos\vartheta$ ($\vartheta$ is the scattering angle)
in $e^{+}e^{-}\to J/\psi\to Y\bar{Y}$ with $Y=\Lambda$, $\Sigma^{+,0}$
and $\Xi^{-,0}$ corresponding to curves in black solid, blue dash-dotted,
green dashed, red dotted and long yellow dash-dotted lines respectively.
The black horizontal line is $\mathcal{D}=0$. }

\end{figure}

The quantum discord $\mathcal{D}[\rho_{Y\bar{Y}}]$ is plotted as
functions of $\cos\vartheta$ for $\Lambda$, $\Sigma^{+,0}$ and
$\Xi^{0,-}$ hyperons in Fig. \ref{fig:discord}. We can see that
the discord of the hyperon-antihyperon system is symmetric with respect
to $\vartheta=\pi/2$ (the transverse scattering angle). In the whole
range of $\vartheta\in[0,\pi]$, the quantum discord of spin states
of $Y\bar{Y}$ is non-zero. Additionally, unlike the steering, the
maximum quantum discord is not necessarily located at $\vartheta=\pi/2$:
for $\Lambda$ and $\Xi^{-,0}$, the discord has two peaks away from
$\vartheta=\pi/2$.


\subsection*{Geometric quantum discord}

Since the quantum discord given in Eq. (\ref{eq:discord}) is generally
hard to evaluate, an alternative geometric measure for the discord
is proposed as \citet{Dakic:2010prl}, 
\begin{equation}
\mathcal{D}_{G}\equiv\min_{\chi\in\Omega_{0}}\left\Vert \rho-\chi\right\Vert ^{2},\label{eq:geometric_distance}
\end{equation}
where $\Omega_{0}$ denotes the set of zero-discord states, $||\rho-\chi||^{2}=\mathrm{Tr}[(\rho-\chi)^{2}]$
is the Hilbert-Schmidt distance between two states, and $\mathcal{D}_{G}$
is known as the \textit{geometric quantum discord}.

\begin{figure}[h]
\includegraphics[scale=0.44]{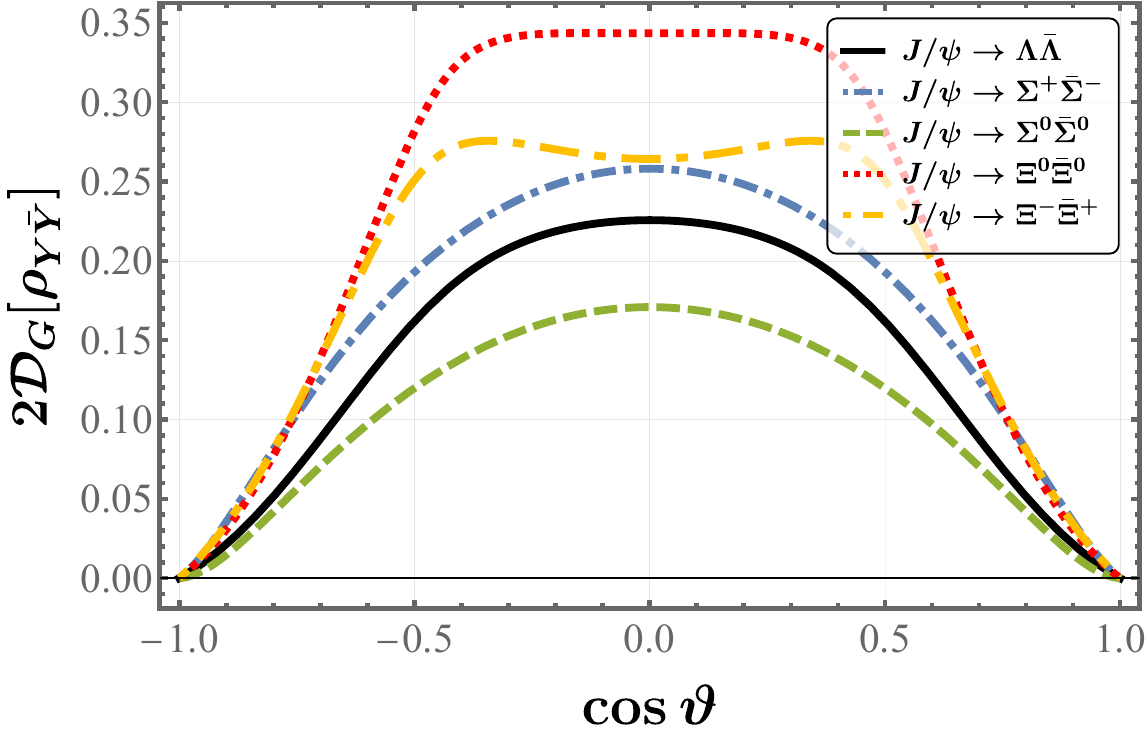}

\caption{\label{fig:geometric_discord}The results for the geometric quantum
discord $2\mathcal{D}_{G}[\rho_{Y\bar{Y}}]$ as functions of $\cos\vartheta$
($\vartheta$ is the scattering angle) in $e^{+}e^{-}\to J/\psi\to Y\bar{Y}$
with $Y=\Lambda$, $\Sigma^{+}$, $\Sigma^{0}$, $\Xi^{0}$ and $\Xi^{-}$
corresponding to black solid, blue dash-dotted, green dashed, red
dotted and long yellow dash-dotted lines respectively. The black horizontal
line is the zero geometric discord $2\mathcal{D}_{G}=0$.}
\end{figure}

The exact and analytical expression for two-qubit states has also
been given in Ref. \citep{Dakic:2010prl}. For the $X$ state given
in Eq. (\ref{eq:X_state}), the geometric discord can be evaluated
as
\begin{equation}
\mathcal{D}_{G}[\rho_{Y\bar{Y}}]=\frac{1}{4}\min\left\{ a^{2}+t_{2}^{2}+t_{3}^{2},t_{1}^{2}+t_{2}^{2}\right\} .\label{eq:geometric}
\end{equation}
It is easy to see that $\mathcal{D}_{G}$ is not normalized to $1$:
its maximum value is $1/2$ for two-qubit states, so it is natural
to consider $2\mathcal{D}_{G}$ as a proper measure in comparison
with the quantum discord $\mathcal{D}$ \citep{Girolami:2011pra}.
The results for $2\mathcal{D}_{G}$ are plotted in Fig. \ref{fig:geometric_discord},
which are non-zero in the full range of the scattering angle and agree
with the original discord in Fig. \ref{fig:discord}. Since $\mathcal{D}_{G}$
is a geometric quantity while $\mathcal{D}$ is an entropy-like quantity,
the relationship between them needs further investigation.

\section{Hierarchy of quantum correlations \label{sec:Hierarchy-of-quantum}}

We can compare the quantum steering and discord with two other quantum
correlations, the entanglement and Bell nonlocality (BN), in hyperon-antihyperon
systems. The latter have been explored in some earlier works \citep{Wu:2024asu,Fabbrichesi:2024rec}.
These different types of quantum correlations in quantum information
theory characterize various aspects of bipartite systems and follow
the hierarchy \citep{Wiseman:2007prl,Adesso:2016jpa,Afik:2022dgh},
\begin{equation}
\mathrm{Bell\ Nonlocality}\subset\mathrm{Steering}\subset\mathrm{Entanglement}\subset\mathrm{Discord}.\label{eq:hierarchy}
\end{equation}


In order to have a fair comparison of different quantum correlations,
we scale the measures of correlations to the range $[0,1]$. Since
the quantum discord is an entropy-like quantity, we can directly use
$\mathscr{D}\equiv\mathcal{D}[\rho_{Y\bar{Y}}]$ in Eq. (\ref{eq:discord}).
We know that the concurrence usually serves as the measure for the
entanglement. However, we adopt in this work the \textit{entanglement
of formation }\citep{Wootters:1998prl} defined as, 
\begin{align}
\mathscr{E} & \equiv h\left(\frac{1+\sqrt{1-\mathcal{C}^{2}[\rho_{Y\bar{Y}}]}}{2}\right)\nonumber \\
 & =h\left(\frac{1+\sqrt{1-t_{2}^{2}}}{2}\right),
\end{align}
where $h(x)$ is the Shannon binary entropy function, and we used
$\mathcal{C}[\rho_{Y\bar{Y}}]=|t_{2}|$ in the second line \citep{Wu:2024asu}.
Obviously, $\mathscr{E}$ is also an entropy-like quantity with $\mathscr{E}\in[0,1]$.


The Bell nonlocality and quantum steering are similar in that they
are quantified by the violation of the Bell and CJWR inequalities
respectively. The bound for the quantum steering is $\mathcal{F}_{3}>1$
and the maximal violation is $\mathcal{F}_{3}^{\max}=\sqrt{3}$. Therefore,
the modified measure of the quantum steering can be given by 
\begin{equation}
\mathscr{S}\equiv\max\left\{ 0,\frac{\mathcal{F}_{3}-1}{\sqrt{3}-1}\right\} \in[0,1].
\end{equation}
Similarly, the bound for the Bell nonlocality is $\mathcal{B}>2$
and the maximum violation is $\mathcal{B}^{\max}=2\sqrt{2}$, the
modified measure of the Bell nonlocality can then be given by 
\begin{equation}
\mathscr{B}\equiv\max\left\{ 0,\frac{\mathcal{B}-2}{2\sqrt{2}-2}\right\} \in[0,1].
\end{equation}


\begin{figure*}
\includegraphics[scale=0.38]{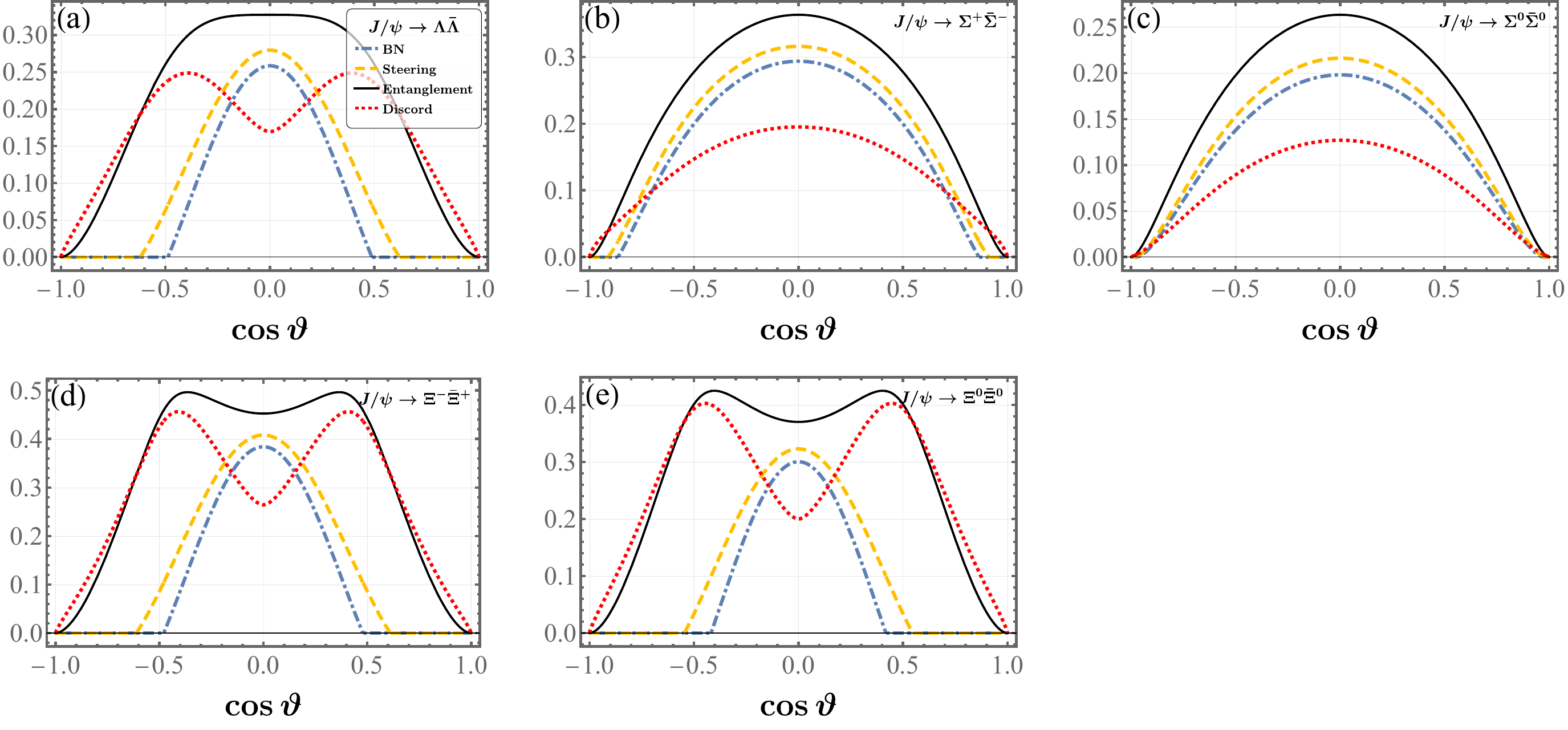}

\caption{\label{fig:hierarchy}Four different types of quantum correlations
as functions $\cos\vartheta$ ($\vartheta$ is the scattering
angle) in $e^{+}e^{-}\to J/\psi\to Y\bar{Y}$. The five panels from
(a) to (e) correspond to $\Lambda$, $\Sigma^{+}$, $\Sigma^{0}$,
$\Xi^{-}$ and $\Xi^{0}$ respectively. The normalized Bell nonlocality
(BN) $\mathscr{B}$ is shown in blue dot-dashed lines, the normalized
steering $\mathscr{S}$ is shown in yellow dashed lines, the normalized
entanglement $\mathscr{E}$ is shown in black solid lines, and the
discord $\mathscr{D}$ is shown in red dotted lines.  }

\end{figure*}


In Fig.~\ref{fig:hierarchy} we compare four quantum correlations,
Bell nonlocality $\mathscr{B}$, steering $\mathscr{S}$, entanglement
$\mathscr{E}$ and discord $\mathscr{D}$, as functions of $\cos\vartheta$
for various hyperon-antihyperon systems in $e^{+}e^{-}\to J/\psi\to Y\bar{Y}$,
with panel (a)-(e) for $\Lambda$, $\Sigma^{+}$, $\Sigma^{0}$, $\Xi^{-}$
and $\Xi^{0}$ respectively. In each panel, $\mathscr{B}$, $\mathscr{S}$,
$\mathscr{E}$ and $\mathscr{D}$ are plotted in the blue dash-dotted,
yellow dashed, black solid, red dotted curves respectively.


\begin{table*}
\caption{\label{tab:hierarchy}The maximal values and ranges for four types
of quantum correlations. The angles $\vartheta_{\mathrm{BN}}^{*}$
and $\vartheta_{\mathrm{steering}}^{*}$ are listed for the non-zero
Bell nonlocality and steering in the range $\left[\vartheta^{*},180^{\circ}-\vartheta^{*}\right]$.
The angle $\vartheta^{\max}$ shown in parentheses corresponds to
the one at which the hyperon-antihyperon system reaches its maximum
entanglement or discord.}

\begin{ruledtabular}
\begin{tabular}{cccccc}
 & $\Lambda\bar{\Lambda}$ & $\Sigma^{+}\bar{\Sigma}^{-}$ & $\Sigma^{0}\bar{\Sigma}^{0}$ & $\Xi^{-}\bar{\Xi}^{+}$ & $\Xi^{0}\bar{\Xi}^{0}$\tabularnewline
\hline 
$\mathscr{B}_{\max}$ & $0.259$ & $0.294$ & $0.198$ & $0.384$ & $0.300$\tabularnewline
$\vartheta_{\textrm{BN}}^{*}$ & $27.93^{\circ}$ & $49.45^{\circ}$ & $14.30^{\circ}$ & $61.38^{\circ}$ & $65.29^{\circ}$\tabularnewline
$\mathscr{S}_{\max}$ & $0.280$ & $0.316$ & $0.216$ & $0.408$ & $0.323$\tabularnewline
$\vartheta_{\textrm{steering}}^{*}$ & $35.70^{\circ}$ & $52.08^{\circ}$ & $10.58^{\circ}$ & $52.41^{\circ}$ & $56.98^{\circ}$\tabularnewline
$\mathscr{E}_{\max}(\vartheta^{\max})$ & $0.327(90^{\circ})$ & $0.363(90^{\circ})$ & $0.263(90^{\circ})$ & $0.496(68.60^{\circ})$ & $0.425(66.26^{\circ})$\tabularnewline
$\mathscr{D}_{\max}(\vartheta^{\max})$ & $0.170(90^{\circ})$ & $0.195(90^{\circ})$ & $0.127(90^{\circ})$ & $0.456(65.86^{\circ})$ & $0.403(63.49^{\circ})$\tabularnewline
\end{tabular}
\end{ruledtabular}

\end{table*}


In Fig.~\ref{fig:hierarchy}, the magnitudes of the Bell nonlocality,
steering and entanglement exhibit a clear hierarchy $\mathscr{B}<\mathscr{S}<\mathscr{E}$,
indicating that the degree of the entanglement is stronger than that
of the steering and than that of the Bell nonlocality. However, the
magnitude of the discord dose not show a clear hierarchical relation
with other three correlations. Even if both the quantum entanglement
and discord are entropy-like quantities, there is no simple dominance
relation between them in magnitudes \citep{Luo:2008pra}, suggesting
that the entanglement and discord represent distinct aspects of quantum
correlations.


Apart from the magnitudes of quantum correlations, by examining the
non-zero regions with respect to the scattering angle $\vartheta$
(or $\cos\vartheta$), we can also explore the hierarchical relation
in them. From Fig. \ref{fig:hierarchy}, it is evident that both the
discord and entanglement exhibit non-zero features in the full range
of the scattering angle (except at two end points $\vartheta=0$ and
$180^{\circ}$). The steerability, however, is restricted to a limited
range centered at the transverse angle angle, and the Bell nonlocality
is even more constrained which lies within the scope of the steering.
Some useful parameters for four correlations are listed in Table \ref{tab:hierarchy}.


In combination of Fig. \ref{fig:hierarchy} and Table \ref{tab:hierarchy},
we see that the Bell nonlocality, steering and entanglement follow
a hierarchical relation, $ \textrm{Bell
Nonlocality} \subset \textrm{Steering} \subset \textrm{Entanglement} \subset \textrm{Discord} $,
indicating that the steering is less restrictive than Bell nonlocality
but more restrictive than the entanglement. Albeit the quantum discord
is proved to be more general and less restrictive than the entanglement,
one can not observe such a relation in hyperon-antihyperon systems,
since both the entanglement and discord are non-zero in the full range
of the scattering angle. The reason is quite simple and direct: the
spin states of hyperon-antihyperon systems are special and show the
subtly between the discord and entanglement. Nevertheless, the rest
part of the hierarchy presented in (\ref{eq:hierarchy}) is confirmed
in hyperon-antihyperon systems.


We also calculate the quantum correlations in $Y\bar{Y}$ systems
through $\psi$(3686) in $e^{+}e^{-}$ annihilation using the available
data for the parameters $\alpha_{\psi}$ and $\Delta\Phi$, see Appendix\@.
\ref{sec:hierarachy_psi2s}.


\section{EMFFs in quantum correlations \label{sec:EMFFs}}

In this section, we discuss the role of electromagnetic form factors
(EMFFs) in quantum correlations in hyperon-antihyperon systems. In
$e^{+}e^{-}\to J/\psi\to Y\bar{Y}$, the $J/\psi$-hyperon vertex
can be written as \citep{Faldt:2017kgy} 
\begin{equation}
\Gamma_{\mu}(p_{1},p_{2})=-ie\left[G_{M}\gamma_{\mu}-\frac{2M}{Q^{2}}\left(G_{M}-G_{E}\right)Q_{\mu}\right],\label{eq:vertex}
\end{equation}
where $P=p_{1}+p_{2}$, $Q=p_{1}-p_{2}$, $M$ is the hyperon mass,
and $G_{E}$ and $G_{M}$ are electric and magnetic form factors of
hyperons as functions $s=P^{2}$. The decay parameters $\alpha_{\psi}$
and $\Delta\Phi$ are defined through $G_{E}$ and $G_{M}$ by 
\begin{eqnarray}
\alpha_{\psi} & = & \frac{s\left|G_{M}\right|^{2}-4M^{2}\left|G_{E}\right|^{2}}{s\left|G_{M}\right|^{2}+4M^{2}\left|G_{E}\right|^{2}},\nonumber \\
G_{E}/G_{M} & = & e^{i\Delta\Phi}\left|G_{E}/G_{M}\right|,\label{eq:phase-ge-gm}
\end{eqnarray}
where $\Delta\Phi$ is the relative phase between two complex-valued
form factors. 

\begin{figure}
\includegraphics[scale=0.44]{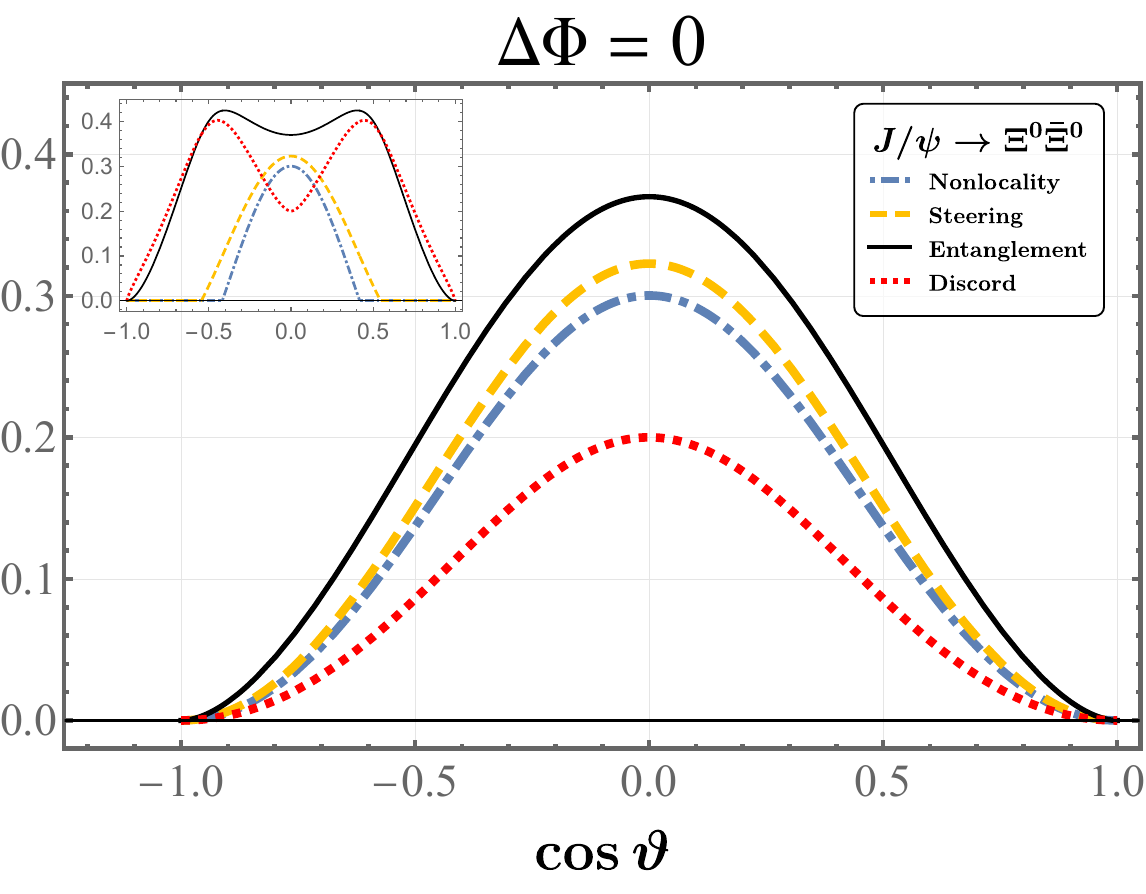}

\caption{\label{fig:zero_phase}The quantum correlations for $\Xi^{0}\bar{\Xi}^{0}$
with $\Delta\Phi=0$. The subfigure shows the result in Fig.~\ref{fig:hierarchy}~(e)
for comparison. The Bell nonlocality $\mathscr{B}$, the steering
$\mathscr{S}$, the entanglement $\mathscr{E}$, and the discord $\mathscr{D}$
are shown in blue dot-dashed, yellow dashed, black solid, and red
dotted lines respectively.}

\end{figure}

It can be seen from Eq. (\ref{eq:at123}) that the quantum correlations
at $\vartheta=90^{\circ}$ only depend on the decay
parameter $\alpha_{\psi}$ and are irrelevant of the phase $\Delta\Phi$,
\begin{align}
\mathscr{D}_{\vartheta=90^{\circ}} & =1-h\left(\frac{1-\alpha_{\psi}}{2}\right),\nonumber \\
\mathscr{E}_{\vartheta=90^{\circ}} & =h\left(\frac{1+\sqrt{1-\alpha_{\psi}^{2}}}{2}\right),\nonumber \\
\mathscr{S}_{\vartheta=90^{\circ}} & =\frac{\sqrt{1+2\alpha_{\psi}^{2}}-1}{\sqrt{3}-1},\nonumber \\
\mathscr{B}_{\vartheta=90^{\circ}} & =\frac{\sqrt{1+\alpha_{\psi}^{2}}-1}{\sqrt{2}-1}.
\end{align}
In order to see the effect of $\Delta\Phi$ in quantum correlations,
we deliberately set $\Delta\Phi$ to zero. In this case, the analytical
expressions of these four types of correlations are reduced to simpler
forms: 
\begin{align}
\mathscr{D}_{\Delta\Phi=0} & =1-h\left(\frac{1+\alpha_{\psi}\cos2\vartheta}{2(1+\alpha_{\psi}\cos^{2}\vartheta)}\right),\nonumber \\
\mathscr{E}_{\Delta\Phi=0} & =h\left(\frac{1}{2}+\frac{\sqrt{(1+\alpha_{\psi})(1+\alpha_{\psi}\cos2\vartheta)}}{2(1+\alpha_{\psi}\cos^{2}\vartheta)}\right),\nonumber \\
\mathscr{S}_{\Delta\Phi=0} & =\frac{\sqrt{1+2\left[\alpha_{\psi}\sin^{2}\vartheta/(1+\alpha_{\psi}\cos^{2}\vartheta)\right]^{2}}-1}{\sqrt{3}-1},\nonumber \\
\mathscr{B}_{\Delta\Phi=0} & =\frac{\sqrt{1+\left[\alpha_{\psi}\sin^{2}\vartheta/(1+\alpha_{\psi}\cos^{2}\vartheta)\right]^{2}}-1}{\sqrt{2}-1}.
\end{align}
Here, we choose $J/\psi\to\Xi^{0}\bar{\Xi}^{0}$ as a representative
to illustrate the effect of $\Delta\Phi$. We plot the correlations
in Fig.~\ref{fig:zero_phase} with and without $\Delta\Phi$. We see
that the profiles of four types of quantum correlations are modified
significantly. With $\Delta\Phi=0$, all correlations are non-zero
in the whole range of the scattering angle except at $\vartheta=0$
and $180^{\circ}$, and the hierarchy becomes $\textrm{Discord} \subset \textrm{Bell
Nonlocality} \subset \textrm{Steering} \subset \textrm{Entanglement}$, which is different
from (\ref{eq:hierarchy}) and the case with non-zero $\Delta\Phi$. 

\section{Quantum correlation as probe to particle's compositeness}

As composite particles, hyperons exhibit spin-correlation patterns that are qualitatively different from those of elementary fermions. In particular, the nonzero relative phase between the EMFFs generates a polarization along the $y$ direction, proportional to $\sin(\Delta\Phi)$ and thus to $\mathrm{Im}(G_{M}G_{E}^{*})$. Such correlations have no analogue in elementary systems such as $t\bar{t}$ pairs~\cite{Afik:2020onf,Fabbrichesi:2021npl,Afik:2022dgh,Afik:2022kwm} or $\tau^{+}\tau^{-}$ leptons~\cite{Fabbrichesi:2022ovb,Ehataht:2023zzt}. The underlying difference arises from the interference between $G_{E}$ and $G_{M}$ in the $J/\psi$-hyperon vertex [Eq.~\eqref{eq:vertex}].

In contrast, $t\bar{t}$ production in $p\bar{p}$ collisions is through two partonic channels, $q\bar{q}\to t\bar{t}$ and $gg\to t\bar{t}$. These channels yield a simple relation between Bell nonlocality and concurrence, $\mathcal{B}=2\sqrt{1+\mathcal{C}^{2}}$, which also appears in $\tau^{+}\tau^{-}$ production~\cite{Wu:2024asu}.

This phenomenological distinction suggests a conjecture: the quantum correlation of particle-antiparticle may serve as a probe to the compositeness of the particle. The complex-valued structure of the hyperon's timelike EMFFs is therefore essential to understand the observed quantum correlation patterns. This motivates a deeper exploration of hyperon-antihyperon systems within the framework of quantum information at colliders such as BEPCII and the proposed STCF~\cite{Achasov:2023gey}.

\section{\label{sec:loophole}Locality Loophole in HEP Experiments}

In the history of testing quantum nonlocality, experimentalists have faced several potential loopholes, including the detection-efficiency loophole~\citep{Pearle:1970prd}, the freedom-of-choice loophole, and the locality loophole~\citep{Bell:1995}. The presence of such loopholes may allow local realism to mimic quantum predictions, thereby undermining the reliability of experimental results. To address these issues, continuous efforts have been made to close these loopholes, culminating in a series of so-called loophole-free Bell tests performed on various experimental platforms~\citep{Hensen:2015ccp,Giustina:2015yza}.

In Refs.~\citep{Fabbrichesi:2021npl,Ehataht:2023zzt,Fabbrichesi:2024rec}, potential loopholes in testing Bell inequalities at high-energy colliders have been discussed. We argue that, since the overall framework and methodology of high-energy collider experiments differ substantially from those of traditional low-energy Bell tests, it is essential to reconsider how such loopholes may arise and manifest in the high-energy context.

In high-energy collider experiments, the spin density matrix of particles is reconstructed via quantum state tomography. The extremely large data samples ($\sim 10^{10}$) provide sufficient statistics and near-perfect detection efficiency, effectively closing the detection-efficiency loophole. Moreover, weak decays, such as $\Lambda \to p \pi^{-}$, serve as intrinsic polarimeters. Since such decays are actually generalized quantum measurements which are fundamentally random and lack any preferred direction, the freedom-of-choice loophole is also excluded. Consequently, the only loophole relevant in the collider context is the locality loophole.

The locality loophole originates from Einstein's special relativity: if the two measurements are timelike separated, one outcome could in principle influence the other via causal signals. To close this loophole, the measurement events must be space-like separated and executed within a sufficiently short time window.


In this section, we consider $e^{+}e^{-} \to \Lambda \bar{\Lambda}$ as an example. Although $\Lambda$ and $\bar{\Lambda}$ hyperons share the same mean lifetime, their individual decay times are not identical. Consequently, if their back-to-back velocities are not sufficiently large, the two decay events may be timelike separated, giving rise to a potential locality loophole. To eliminate this loophole in high-energy collider experiments, one must select events in which the two decays are spacelike separated and reject those that are timelike separated. This selection can be implemented by applying cuts on the decay times.  

\begin{figure*}
\centering
\includegraphics[scale=0.7]{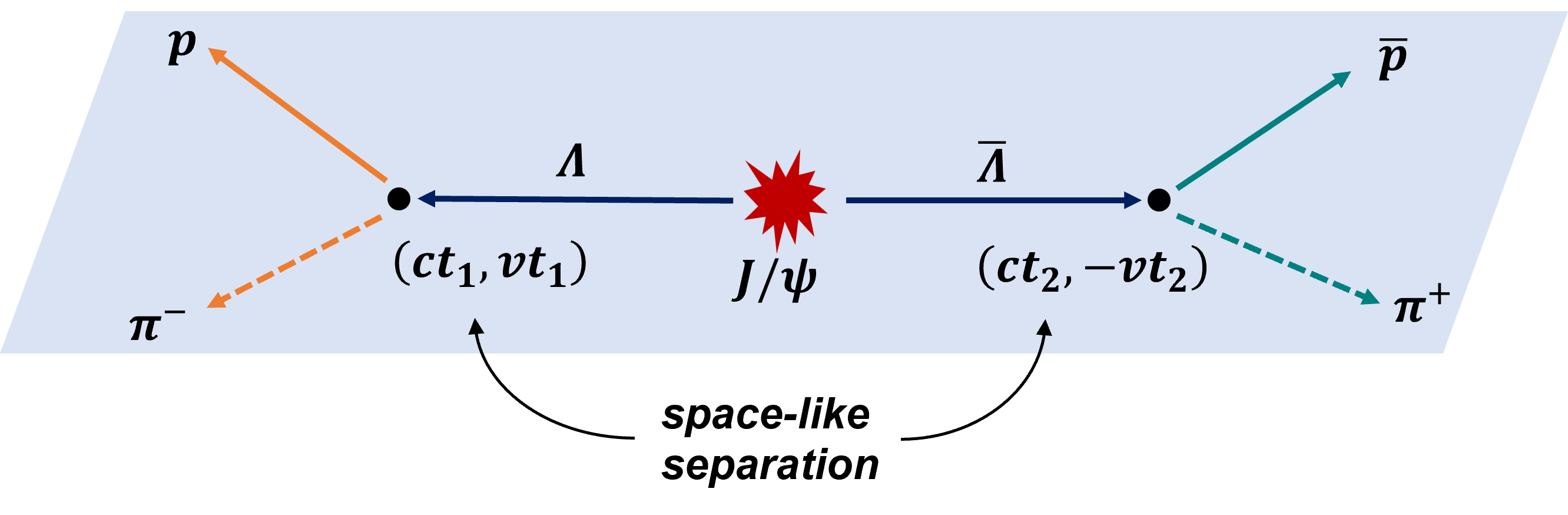}
\caption{Schematic diagram of the decay process $J/\psi \to \Lambda \bar{\Lambda} \to p \pi^{+} \bar{p} \pi^{-}$. In orer to close the locality loophole, two space-time corrodinates $(ct_1, vt_1)$ and $(ct_2, -vt_2)$ should be space-like separation.}
\label{fig:loophole}
\end{figure*}

In the process $e^{+}e^{-} \to J/\psi \to \Lambda \bar{\Lambda}$, the center-of-mass energy is fixed at $\sqrt{s} = 3.096~\mathrm{GeV}$, yielding a $\Lambda$ velocity of approximately $0.69c$. Suppose the hyperon and antihyperon propagate back-to-back and undergo weak decays at times $t_{1}$ and $t_{2}$, respectively. Since the decay acts as a generalized measurement, the corresponding spacetime coordinates can be written as $(ct_{1}, vt_{1})$ and $(ct_{2}, -vt_{2})$. The condition for spacelike separation requires  
\begin{equation}
\Delta s^{2} = c^{2}(t_{1}-t_{2})^{2} - v^{2}(t_{1}+t_{2})^{2} < 0,
\end{equation}  
as illustrated in Fig.~\ref{fig:loophole}. This inequality leads to  \begin{equation}
\frac{|t_{1}-t_{2}|}{t_{1}+t_{2}} < \beta_{\Lambda},
\label{eq:space}
\end{equation}  
where $\beta_{\Lambda} = v/c = \sqrt{1 - 4M^{2}/s}$ is $\Lambda$'s velocity in the unit of light speed. Equation~\eqref{eq:space} shows that spacelike separation is satisfied whenever the time difference between two decays is sufficiently small compared to their average lifetime. For $J/\psi \to \Lambda \bar{\Lambda}$, one finds $\beta_{\Lambda} \simeq 0.69$, implying that the ratio $|t_{1}-t_{2}|/(t_{1}+t_{2})$ must be less than 0.69 to ensure spacelike separation.

An unstable particle's lifetime follows an exponential distribution $f(t) = (1/T) e^{-t/T}$, where $T$ is the mean lifetime. It is well accepted that the decay times $t_{1}$ and $t_{2}$ for $\Lambda$ and $\bar{\Lambda}$ are statistically independent (actually this hypothesis can be tested in experiments by measuring the lifetime correlation in hyperon-antihyperon systems~\citep{Tang:2025oav}). From probability theory, it follows that the random variable appearing on the left-hand side of Eq.~\eqref{eq:space} is uniformly distributed over the interval $[0,1]$: 
\begin{equation}
    \frac{|t_{1}-t_{2}|}{t_{1}+t_{2}} \sim \text{Uniform}[0,1].
    \label{eq:uniform}
\end{equation}
Consequently, the probability for spacelike separation is determined solely by the hyperon velocity, independent of $T$:  
\begin{equation}
    \Pr(\Delta s^{2} < 0) = \beta_{\Lambda}.
\end{equation}
This result can be generalized to other hyperon pairs. 

\begin{figure}[htbp]
\centering
\includegraphics[scale=0.7]{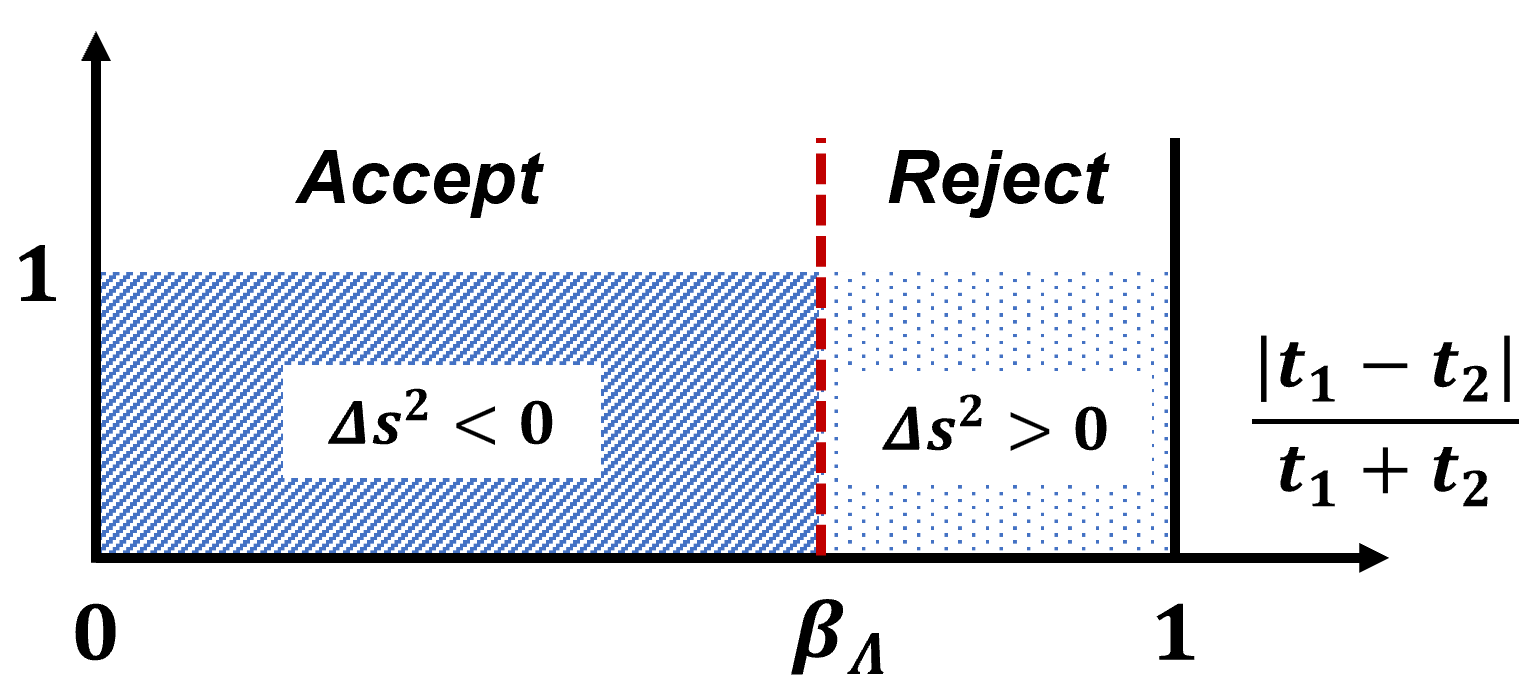}
\caption{The fraction of events in which $\Lambda$ and $\bar{\Lambda}$ decays are space-like separated is given by $\beta_{\Lambda}$. The random variable in the left-hand side of Eq.~(\ref{eq:space}) is a uniform distribution in $[0, 1]$.
}
\label{fig:uniform}
\end{figure}

In collider experiments, selecting events that satisfy Eq.~\eqref{eq:space} effectively closes the locality loophole. Combined with Eq.~\eqref{eq:uniform}, we see that the selection condition in Eq.~\eqref{eq:space} is independent of the mean lifetime and depends only on the particle's velocity. In $e^{+}e^{-} \to J/\psi \to \Lambda \bar{\Lambda}$, the final-state hyperons have $\beta_{\Lambda} \simeq 0.69$, meaning that approximately $69\%$ of all events are space-like separated and accepted, or $31\%$ of events are time-like separated and then rejected (see Fig.~\ref{fig:uniform}). The criterion in Eq.~\eqref{eq:space} implies that increasing the number of accepted events requires larger $\beta_{\Lambda}$ (or larger $\sqrt{s}$ equivalently). Therefore, in practice, it is preferable to use data at higher center-of-mass energies for a given hyperon; for instance, choosing $\psi(3686)$ rather than $J/\psi$. Of course, experimental analyses must also take into account the size difference in datasets. At BESIII, for example, the $J/\psi$ dataset contains $10 \times 10^{9}$ events, whereas the $\psi(3686)$ dataset contains $2.7 \times 10^{9}$ events. 

As a final remark, we emphasize the distinction between \emph{testing quantum nonlocality} and \emph{measuring nonlocal witnesses}. The former aims to verify the nonlocal nature of quantum mechanics at a high-energy scale. In such experiments, it is essential to carefully rule out the locality loophole to ensure that the results are trustworthy~\cite{CMS:2024zkc,BESIII:2025vsr}. The latter, in contrast, focuses on quantifying quantum correlations in high-energy systems, where the notion of loopholes does not directly apply.

The existence and relevance of loopholes in collider-based tests remain under active discussion within the high-energy physics community; see, for example, Refs.~\cite{Bechtle:2025ugc,Abel:2025skj,Fabbrichesi:2025psr,Low:2025aqq,Demina:2024dst,Fabbrichesi:2025aqp}.


\section{\label{sec:decoherence}Decoherence and Quantum Correlations}

Quantum decoherence plays a vital role in understanding the quantum nature of physical systems, characterizing their evolution under the influence of environmental interactions~\cite{Schlosshauer:2019ewh}. Recent studies have introduced decoherence effects into the description of high-energy processes~\cite{Aoude:2025ovu,Gu:2025ijz}. At collider experiments such as BESIII, the observation of quantum correlations may be challenged by decoherence phenomena~\cite{Caban:2014pra,Fabbrichesi:2024rec}, which arise from interactions between high-energy particles and the detector materials~\cite{BESIII:2023clq,BESIII:2023trh,BESIII:2024geh,Dai:2024cpc,Demina:2024dst}. In this section, we discuss the role of quantum decoherence in detectors and its potential impact on the measurement of quantum correlations.

 
We take the decay $\eta_c \to \Lambda\bar{\Lambda}$ as an example. According to its mean lifetime $\sim 2.6\times 10^{-10}$ s and $\beta_\Lambda \approx 0.66$, $\Lambda$ will undergo a secondary decay at an average distance of $5.1$ cm from the production vertex. Given that the inner radius of the beam pipe at BESIII is 3.15 cm, it is reasonable to assume that most $\Lambda$'s interact with the beam pipe wall and potentially enter the first few layers of the Main Drift Chamber (MDC).
When $\Lambda$ hyperons enter the MDC, their spatial degrees of freedom are modified due to the localization of their trajectories, and their spin orientations may also change. 


Owing to the exponential distribution of their lifetimes, a fraction of short-lived $\Lambda$ hyperons decay inside the beam pipe. We exclude these hyperons from our model, as they never interact with any detector materials. In contrast, the longer-lived $\Lambda$ hyperons traverse the beam pipe or other detector components and interact with detector materials. The latter $\Lambda$'s are the focus of our analysis, and we will describe their spin evolution within the framework of open quantum systems. 


In quantum information theory, the evolution of an open quantum system can be described by the Lindblad (or Gorini-Kossakowski-Sudarshan-Lindblad) master equation~\cite{Lindblad:1976generators,Gorini:1975nb,Breuer:2002pc}:
\begin{equation}
\frac{\mathrm{d}\rho}{\mathrm{d} t} = -i\left[\mathbf{H},\rho\right] +\sum_{\ell}\Big(\mathbf{L}_{\ell}\rho\mathbf{L}_{\ell}^{\dagger}-\frac{1}{2}\{\mathbf{L}_{\ell}^{\dagger}\mathbf{L}_{\ell},\rho\}\Big),
\label{eq:Lindblad}
\end{equation}
where $\rho$ is the spin density operator of a particle. 
Equation~(\ref{eq:Lindblad}) contains both the unitary dynamics generated by the effective Hamiltonian $\mathbf{H}$ and the non-unitary dynamics induced by system-environment interactions, represented by the jump operators $\mathbf{L}_\ell$.


In principle, the explicit form of the jump operators $\mathbf{L}_{\ell}$ should be derived from a microscopic model of hyperon-material interactions. However, for the purposes of this work, we adopt a phenomenological approach. Specifically, we model the decoherence of the $\Lambda$ hyperon via the phase damping channel:
\begin{equation}
\frac{\mathrm{d}\rho_{\Lambda}}{\mathrm{d}t}= -\frac{\gamma}{2}[\sigma_{z},[\sigma_{z},\rho_{\Lambda}]],
\label{eq:dephasing_1}
\end{equation}
where we omit the unitary term $-i[\mathbf{H}, \rho_{\Lambda}]$ since it does not affect quantum correlations, and take $\sqrt{\gamma}\sigma_z$ as the sole jump operator.


Here, the $z$ axis is defined along the momentum (flight) direction of the $\Lambda$ hyperon. As the $\Lambda$ traverses the beam pipe or detector materials, it may undergo weak, stochastic interactions with the surrounding medium. These processes typically act as repeated partial 'measurements' of their spin projection along the momentum direction. From the perspective of open quantum systems, such interactions preferentially select the eigenbasis of $\sigma_z$ as the pointer basis - the basis that remains stable under environmental monitoring. This leads to progressive suppression of coherence between spin-up and spin-down states along the flight direction, while populations in that basis remain unaffected. Mathematically, this behavior is captured by Eq.~(\ref{eq:dephasing_1}), which describes a continuous weak measurement of the spin-$z$ component, consistent with the expected symmetry in the rest frame of the hyperon~\cite{Jacobs:2006cp}.


The phase damping channel in Eq.~\eqref{eq:dephasing_1} can be equivalently written in the operator-sum (Kraus) reperesentation: 
\begin{equation}
\rho_{\Lambda}\longmapsto\mathcal{E}(\rho_{\Lambda})\equiv\mathbf{K}_{0}\rho_{\Lambda}\mathbf{K}_{0}+\mathbf{K}_{1}\rho_{\Lambda}\mathbf{K}_{1},
\label{eq:dephasing}
\end{equation}
where the Kraus operators are:
\begin{equation}
\mathbf{K}_{0}=\begin{bmatrix}1 & 0\\
0 & \sqrt{1-\zeta}
\end{bmatrix},\qquad\mathbf{K}_{1}=\begin{bmatrix}0 & 0\\
0 & \sqrt{\zeta}
\end{bmatrix},
\end{equation}
with $\sqrt{1-\zeta} \equiv e^{-2 \gamma t}$. Acting on the Bloch vector $\mathbf{B^{+}}=(B_x^{+},B^{+}_y,B^{+}_z)$, the channel yields
\begin{equation}
(B_{x}^{+},B_{y}^{+},B_{z}^{+})\longmapsto(e^{-2 \gamma t}B_{x}^{+}, e^{-2 \gamma t}B_{y}^{+},B_{z}^{+}).
\label{eq:relaxing}
\end{equation}
which is the solution of the master equation \eqref{eq:dephasing_1}: transverse components decay with the rate $2\gamma$ while the longitudinal component is conserved. In NMR terminology this corresponds to $T_2$-type relaxation with $1/T_{2} = 2\gamma$. Phase damping is commonly used to model the loss of coherence (information) induced by random scattering processes, analogous to the interactions experienced by a $\Lambda$ hyperon traversing detector materials~\cite{Nielsen2015}.


For a two-qubit system such as $\Lambda\bar{\Lambda}$, we assume that the antiparticle undergoes the same decoherence process as the particle [Eqs.~\eqref{eq:dephasing_1},~\eqref{eq:dephasing}]. The complete two-qubit channel is then given by $\rho_{\Lambda\bar{\Lambda}} \mapsto \tilde{\rho}_{\Lambda \bar{\Lambda}} \equiv \mathcal{E}\otimes\mathcal{E}\!\left(\rho_{\Lambda\bar{\Lambda}}\right)$. Since decoherence corresponds to quantum information leaking into the environment, it reduces the correlation between $\Lambda$ and $\bar{\Lambda}$. 

\begin{figure}[htbp]
\centering
\includegraphics[scale=0.54]{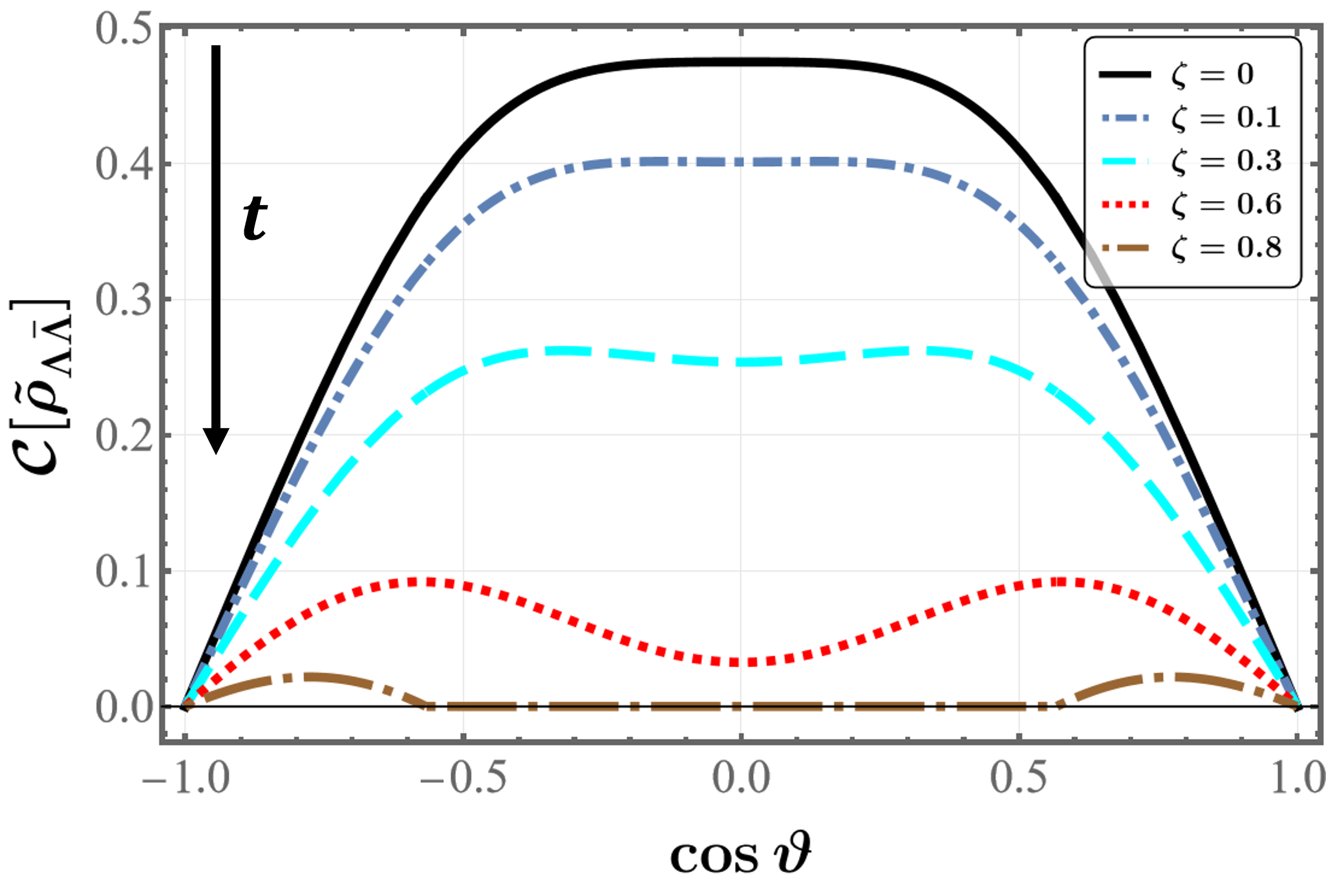}
\caption{The entanglement between $\Lambda$ and $\bar{\Lambda}$ will decrease due to the quantum decoherence effect. The decoherence time $t$ is related to the factor $\zeta$ as $\sqrt{1-\zeta} = e^{-2\gamma t}$.}
\label{fig:decoherence}
\end{figure}

We can employ the concurrence as a quantitative measure to illustrate how decoherence degrades the entanglement of the $\Lambda\bar{\Lambda}$ system. The numerical results are presented in Fig.~\ref{fig:decoherence}. 
We observe that the concurrence decreases monotonically with the decoherence parameter $\zeta$ and eventually vanishes in the strong-decoherence limit, reflecting the complete loss of the entanglement between $\Lambda$ and $\bar{\Lambda}$.


In measurements of the process $e^+e^- \to \Lambda\bar{\Lambda}$, the angular distributions of the secondary decays are used to extract the parameter $\Delta\Phi$ without accounting for decoherence effects. This implies that the current experimental results (e.g., $\Delta\Phi = 0.752$ for $\Lambda$) already include such effects, leading to an underestimation of the entanglement relative to its value at the production vertex.


To obtain more accurate values of the quantum correlation and also $\Delta\Phi$, a promising approach is proposed in Ref.~\citep{Fabbrichesi:2024rec}: using $\eta_c/\chi_{c0} \to \Lambda\bar{\Lambda}$ to benchmark the decoherence effect for the $\Lambda$ hyperon. The above decay channels produce two Bell states by symmetry:
\begin{equation}
|\Psi^{\pm}\rangle = \frac{1}{\sqrt{2}} \left( |\uparrow \downarrow \rangle \pm |\downarrow \uparrow \rangle \right).
\label{eq:Bellpm}
\end{equation}
By comparing the theoretical predictions in Eq.~(\ref{eq:Bellpm}) with experimental results reconstructed via quantum state tomography, one can directly confirm the presence of decoherence for the $\Lambda$ hyperon in the detector and identify the specific decoherence channel, such as determining the relaxation time $T_2$ in a dephasing channel. 


Once the form of the quantum channel, such as $T_{2}$, is established from processes like $\eta_{c}/\chi_{c0}\to\Lambda\bar{\Lambda}$, the original values of $B_{y}^{\pm}$ (without decoherence) can be extracted from experimental measurements of the polarization after relaxation using Eq.~\eqref{eq:relaxing}. Therefore, the experimental value of $\Delta\Phi$ in $J/\psi \to \Lambda\bar{\Lambda}$ at BESIII -- currently extracted without accounting for decoherence -- can be corrected accordingly. This procedure would make it possible to recover the intrinsic quantum correlations in $\Lambda$ and $\bar{\Lambda}$, free from the suppression caused by quantum decoherence.


\section{Summary and outlook \label{sec:Summary} }

We investigated two types of quantum correlations, the quantum steering
and discord, in hyperon-antihyperon systems in $e^{+}e^{-}\to J/\psi\rightarrow Y\bar{Y}$,
where $Y$ represents octet spin-1/2 hyperons in the ground state:
$\Lambda$, $\Sigma^{+}$, $\Sigma^{0}$, $\Xi^{-}$ and $\Xi^{0}$. 
The analytical expressions were derived for the quantum steering and
discord based on the $X$-shaped density operator for $Y\bar{Y}$
systems. The results depend on the parameters $\alpha_{\psi}$ and
$\Delta\Phi$ measured in BESIII experiments. 
We found that the steerability of certain hyperon-antihyperon
system is non-zero within a range centered around the scattering angle
$\vartheta=90^{\circ}$, while the discord is non-zero at all scattering
angles. This aligns with the fact that almost all quantum states have
non-zero discord \citep{Ferraro:2010pra}.
We partially confirmed the hierarchy relations among four types of quantum correlations.
We examined how electromagnetic form factors influence the correlations
in hyperon-antihyperon systems. 


The locality loophole and detector-induced decoherence are two major concerns in collider-based tests of quantum correlations. We propose a method to address the locality loophole, which may provide a path to ruling out local realism in high-energy collisions. In addition, we build a phenomenological model based on the Lindblad master equation to describe quantum decoherence. This framework offers insight into the decoherence mechanism of hyperons and suggests a strategy for extracting more accurate values of some spin correlation parameters in $J/\psi \to \Lambda\bar{\Lambda}$ from experimental data. 


Recent preliminary measurements indicate an oscillatory behavior in $\lvert G_{E}/G_{M}\rvert$ and in $\Delta\Phi$ as functions of the collision energy $\sqrt{s}$, and several theoretical models have been proposed to account for this phenomenon~\cite{Yang:2019mzq,Haidenbauer:2020wyp,Wan:2021ncg,Chen:2023oqs,Yan:2023nlb,Chen:2024luh,Wang:2024qbl}. This oscillatory pattern may reflect nontrivial spin correlations in the $Y\bar{Y}$ system. With increasing data statistics and further theoretical development, it may become feasible to systematically investigate the energy dependence of quantum correlations in $Y\bar{Y}$ production across the continuum region of $e^{+}e^{-}$ annihilation, beyond the vicinity of specific resonances. 

Quantum properties such as Bell nonlocality, steering, entanglement, and discord may in turn serve as sensitive probes to hadrons' structures such as EMFFs and to fundamental symmetries in their interactions~\cite{Du:2024sly}. The distinctive quantum features of hyperon-antihyperon systems also raise the possibility of using quantum correlations to test whether a particle is composite or elementary. 


From the perspective of high-energy physics, this work, together with
our previous study \citep{Wu:2024asu}, presents a comprehensive picture
for the quantum correlations in hyperon-antihyperon systems in $e^{+}e^{-}$
annihilation. The introduction of the quantum steering and discord
provides new witness for quantum correlations in high-energy physics.
From the quantum information point view, our work advances the potential
for using high-energy particle colliders to study quantum information,
expanding the territory for testing the foundation of quantum mechanics
across all energy scales. Furthermore, studying quantum correlations
in hyperon systems provides new insights into the properties of spin,
a continuously active field of research \citep{Czachor:1997pra,Gingrich:2002prl,Peres:2004rmp,Kurashvili:2022ybg,Rembielinski:2019qbe}.


\begin{acknowledgments}
S. Wu acknowledges helpful discussions with Y. Du during his visit to 
Institute of Modern Physics, Chinese Academy of Sciences. 
This work is supported by the National Natural Science Foundation
of China (NSFC) under Grant Nos.~12305010 and 12135011.
\end{acknowledgments}

\appendix

\section{Detecting steerability via entanglement \label{sec:steering_en}}

Apart from the steering inequality, there is another method to quantify
the steering based on the quantum entanglement \citep{Das:2019pra,Zhang:2021pra}.

Considering a two-qubit state, the steering from Bob to Alice can
be witnessed if the density matrix $\rho_{A\shortleftarrow B}$ is
defined as
\begin{equation}
\rho_{A\shortleftarrow B}=\frac{1}{\sqrt{3}}\rho_{AB}+\left(1-\frac{1}{\sqrt{3}}\right)\rho_{A}\otimes\frac{1}{2},
\end{equation}
is entangled. If the original $\rho_{AB}$ is a $X$ state, the steered
state $\rho_{A\shortleftarrow B}$ also preserves the $X$-structure
\begin{equation}
\rho_{A\shortleftarrow B}=\begin{bmatrix}\frac{1}{\sqrt{3}}\rho_{11}+r & 0 & 0 & \frac{1}{\sqrt{3}}\rho_{14}\\
0 & \frac{1}{\sqrt{3}}\rho_{22}+r & \frac{1}{\sqrt{3}}\rho_{23} & 0\\
0 & \frac{1}{\sqrt{3}}\rho_{23} & \frac{1}{\sqrt{3}}\rho_{33}+s & 0\\
\frac{1}{\sqrt{3}}\rho_{14} & 0 & 0 & \frac{1}{\sqrt{3}}\rho_{44}+s
\end{bmatrix},
\end{equation}
where $r=\frac{3-\sqrt{3}}{6}(\rho_{11}+\rho_{22})$ and $s=\frac{3-\sqrt{3}}{6}(\rho_{33}+\rho_{44})$.
Form the Peres-Horodecki entanglement criterion (also from the Wootters'
concurrence), $\rho_{A\shortleftarrow B}$ is entangled iff the concurrence
is greater than zero, i.e. $\mathcal{C}[\rho_{A\shortleftarrow B}]>0$.
In other words, $\rho_{AB}$ is steerable from Bob to Alice if the
concurrence of steered state $\mathcal{C}[\rho_{A\shortleftarrow B}]>0$,
with 
\begin{align}
\mathcal{C}[\rho_{Y\shortleftarrow\bar{Y}}] & =\frac{1}{2\sqrt{3}}\max\biggl\{0,\nonumber \\
 & \left|t_{1}-t_{2}\right|-\sqrt{(\sqrt{3}-t_{3})^{2}-(\sqrt{3}-1)^{2}a^{2}},\nonumber \\
 & \left|t_{1}+t_{2}\right|-\sqrt{(\sqrt{3}+t_{3})^{2}-(\sqrt{3}+1)^{2}a^{2}}\biggr\}.
\end{align}

\begin{figure}[h]
\includegraphics[scale=0.44]{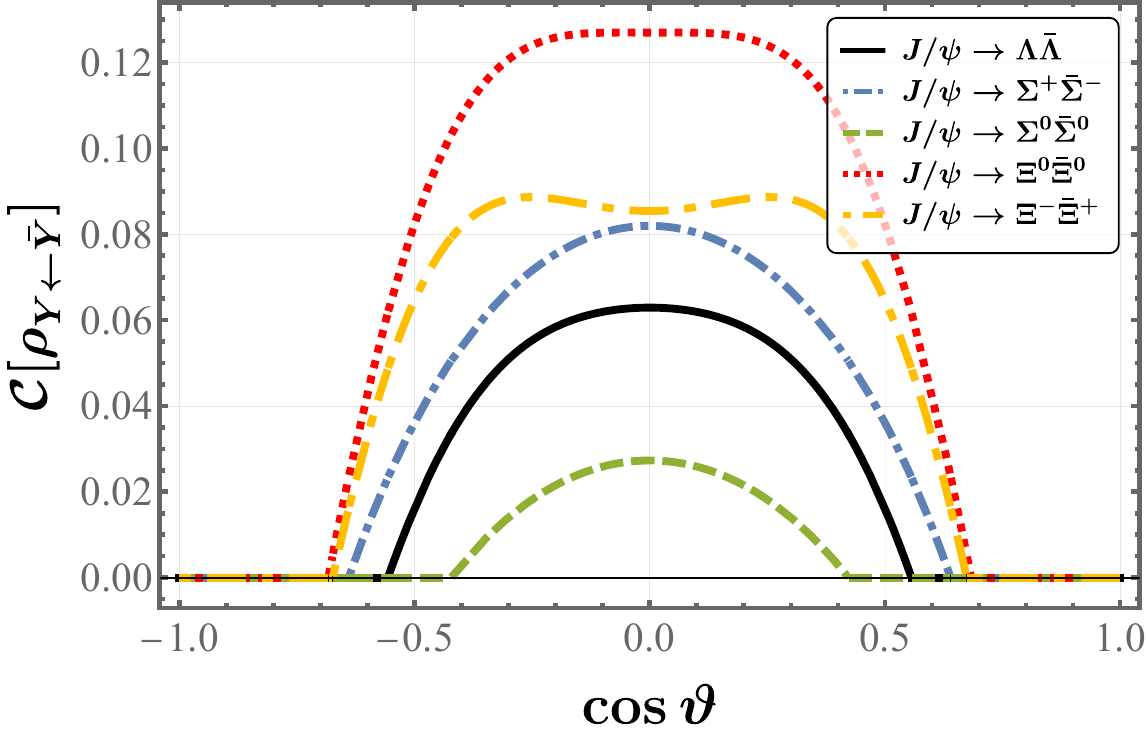}

\caption{\label{fig:Concurrence}The results for the concurrence $\mathcal{C}[\rho_{Y\shortleftarrow\bar{Y}}]$
as functions of $\cos\vartheta$ ($\vartheta$ is the scattering angle)
in $e^{+}e^{-}\to J/\psi\to Y\bar{Y}$ with $Y=\Lambda$, $\Sigma^{+}$,
$\Sigma^{0}$, $\Xi^{0}$ and $\Xi^{-}$ in the black solid, blue
dash-dotted, green dashed, red dotted and long yellow dash-dotted
lines respectively. }
\end{figure}

The results for $\mathcal{C}[\rho_{Y\shortleftarrow\bar{Y}}]$ are
shown in Fig. \ref{fig:Concurrence}. We can see that detecting steerablity
by entanglement has a similar result compared with CJWR inequality
--- the $Y\bar{Y}$ state can be steered within the range around
the transverse scattering angle. However, some details are different.
But the entanglement criterion is also a sufficient but not necessary
condition for the quantum steering.


\section{Quantum correlations in $Y\bar{Y}$ systems through $\psi(3686)$
\label{sec:hierarachy_psi2s}}

\begin{figure*}
\includegraphics[scale=0.38]{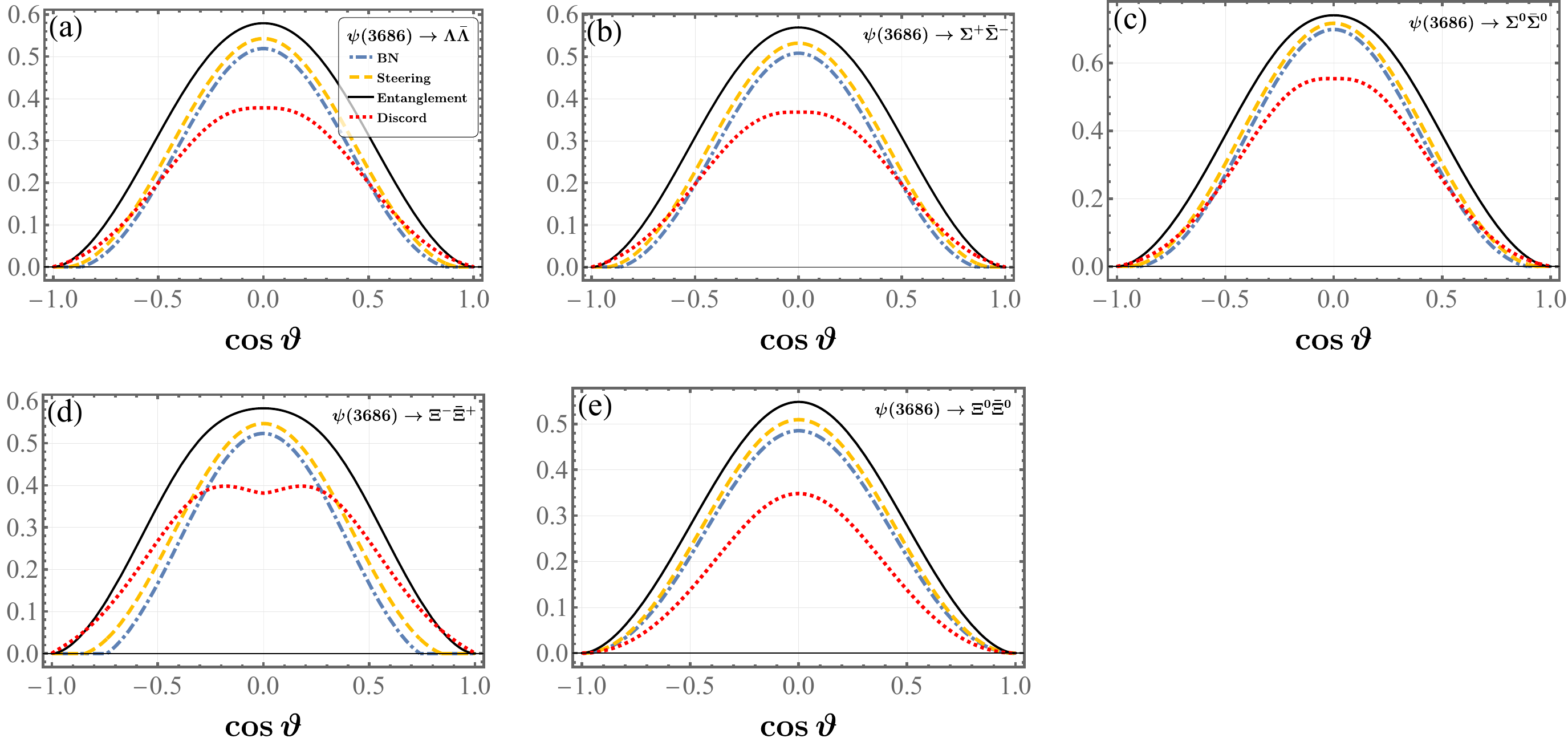}

\caption{\label{fig:hierarachy_psi2s}Four types of quantum correlations as
functions $\cos\vartheta$ in $e^{+}e^{-}\rightarrow\psi(3686)\rightarrow Y\bar{Y}$.
The panels from (a) to (e) correspond to $\Lambda$, $\Sigma^{+}$,
$\Sigma^{0}$, $\Xi^{-}$ and $\Xi^{0}$ respectively. In each panel,
the results for the Bell nonlocality $\mathscr{B}$, steering $\mathscr{S}$,
entanglement $\mathscr{E}$ and discord $\mathscr{D}$ are shown in
blue dot-dashed, yellow dashed, black solid and red dotted lines.}
\end{figure*}

In this appemdix, we calculate the quantum correlations in $Y\bar{Y}$
systems through $\psi$(3686) in $e^{+}e^{-}$ annihilation, using
the experimental data for the parameters $\alpha_{\psi}$ and $\Delta\Phi$
in Table \ref{tab:decay_parameters_psi2s}. The results are shown
in Fig. \ref{fig:hierarachy_psi2s}. This section serves as a complementary
to Sec. \ref{sec:Hierarchy-of-quantum} and aims to extend the investigation
from $J/\psi$ to $\psi(3686)$.


\begin{table*}
\caption{\label{tab:decay_parameters_psi2s}Some parameters in $e^{+}e^{-}\rightarrow\psi(3686)\rightarrow Y\bar{Y}$,
where $Y\bar{Y}$ is a pair of ground-state octet hyperons.}

\begin{ruledtabular}
\begin{tabular}{ccccc}
 & Branching ratio ($\times10^{-4}$) & $\alpha_{\psi}$ & $\Delta\Phi/\mathrm{rad}$ & Ref\tabularnewline
\hline 
$\psi(3686)\to\Lambda\bar{\Lambda}$ & $3.81\pm0.13$ & $0.69\pm0.07\pm0.02$ & $0.40\pm0.14\pm0.03$ & \citep{ParticleDataGroup:2022pth,BESIII:2023euh}\tabularnewline
$\psi(3686)\to\Sigma^{+}\bar{\Sigma}^{-}$ & $2.82\pm0.09$ & $0.682\pm0.030\pm0.011$ & $0.397\pm0.07\pm0.014$ & \citep{ParticleDataGroup:2022pth,BESIII:2020fqg}\tabularnewline
$\psi(3686)\to\Sigma^{0}\bar{\Sigma}^{0}$ & $2.35\pm0.09$ & $0.814\pm0.028\pm0.028$ & $0.512\pm0.085\pm0.034$ & \citep{ParticleDataGroup:2022pth,BESIII:2024nif}\tabularnewline
$\psi(3686)\to\Xi^{-}\bar{\Xi}^{+}$ & $2.87\pm0.11$ & $0.693\pm0.048\pm0.049$ & $0.667\pm0.111\pm0.058$ & \citep{ParticleDataGroup:2022pth,BESIII:2022lsz}\tabularnewline
$\psi(3686)\to\Xi^{0}\bar{\Xi}^{0}$ & $2.3\pm0.4$ & $0.665\pm0.086\pm0.081$ & $-0.050\pm0.150\pm0.020$ & \citep{ParticleDataGroup:2022pth,BESIII:2023drj}\tabularnewline
\end{tabular}
\end{ruledtabular}

\end{table*}


\bibliographystyle{apsrev4-2}
\bibliography{refs}

@article{ParticleDataGroup:2022pth,
  author = {Workman, R. L. and others},
    collaboration = {Particle Data Group},
  title = {Review of Particle Physics},
  journal = {Progress of Theoretical and Experimental Physics},
  volume = {2022},
  pages = {083C01},
  year = {2022},
  doi = {10.1093/ptep/ptac097}
}

@article{Afik:2022dgh,
  title = {Quantum Discord and Steering in Top Quarks at the LHC},
  author = {Afik, Yoav and de Nova, Juan Ram\'on Mu\~noz},
  journal = {Phys. Rev. Lett.},
  volume = {130},
  issue = {22},
  pages = {221801},
  numpages = {6},
  year = {2023},
  month = {May},
  publisher = {American Physical Society},
  doi = {10.1103/PhysRevLett.130.221801},
  url = {https://link.aps.org/doi/10.1103/PhysRevLett.130.221801}
}

@article{Afik:2020onf,
  title={Entanglement and quantum tomography with top quarks at the LHC},
  author={Y. Afik and J. R. M. de Nova},
  journal={Eur. Phys. J. P},
  volume={136},
  pages={907},
  year={2021},
  publisher={Springer},
  url={https://link.springer.com/article/10.1140/epjp/s13360-021-01902-1}
}

@article{Afik:2022kwm,
   title={Quantum information with top quarks in QCD},
   volume={6},
   doi={10.22331/q-2022-09-29-820},
   journal={Quantum},
   author={Y. Afik and J. R. M. de Nova},
   year={2022},
   pages={820},
   url={http://dx.doi.org/10.22331/q-2022-09-29-820}
}

@article{Fabbrichesi:2021npl,
  title = {Testing Bell Inequalities at the LHC with Top-Quark Pairs},
  author = {M. Fabbrichesi and R. Floreanini and G. Panizzo},
  journal = {Phys. Rev. Lett.},
  volume = {127},
  issue = {16},
  pages = {161801},
  numpages = {5},
  year = {2021},
  publisher = {American Physical Society},
  doi = {10.1103/PhysRevLett.127.161801},
  url = {https://link.aps.org/doi/10.1103/PhysRevLett.127.161801}
}

@article{BESIII:2018cnd,
         title = {Polarization and entanglement in baryon–antibaryon pair production in electron–positron annihilation},
	volume = {15},
	issn = {1745-2481},
	url = {https://doi.org/10.1038/s41567-019-0494-8},
	doi = {10.1038/s41567-019-0494-8},
	number = {7},
	journal = {Nature Physics},
	author = {Ablikim, Medina and  others},
	collaboration = {BESIII Collaboration},
	month = jul,
	year = {2019},
	pages = {631--634},
}

@article{Qian:2020ini,
  title = {Nonlocal correlation of spin in high energy physics},
  author = {C. Qian and J.-L. Li and A. S. Khan and C.-F. Qiao},
  journal = {Phys. Rev. D},
  volume = {101},
  issue = {11},
  pages = {116004},
  numpages = {5},
  year = {2020},
  publisher = {American Physical Society},
  doi = {10.1103/PhysRevD.101.116004},
  url = {https://link.aps.org/doi/10.1103/PhysRevD.101.116004}
}

@article{Tornqvist:1980af,
  title={Suggestion for {Einstein-Podolsky-Rosen} experiments using reactions like $e^+ e^-\to\symup{\Lambda}\bar{\symup{\Lambda}}\to p\pi^{-}\bar{p}\pi^{+}$},
  author={N. A. T{\"o}rnqvist},
  journal={Found. Phys.},
  volume={11},
  pages={171},
  year={1981},
  publisher={Springer},
  url={https://link.springer.com/article/10.1007/BF00715204}
}

@article{Clauser:1974tg,
  title = {Experimental consequences of objective local theories},
  author = {J. F. Clauser and M. A. Horne},
  journal = {Phys. Rev. D},
  volume = {10},
  issue = {2},
  pages = {526},
  year = {1974},
  publisher = {American Physical Society},
  doi = {10.1103/PhysRevD.10.526},
  url = {https://link.aps.org/doi/10.1103/PhysRevD.10.526}
}

@article{Aspect:1981zz,
  title = {Experimental Tests of Realistic Local Theories via Bell's Theorem},
  author = {A. Aspect and P. Grangier and G. Roger},
  journal = {Phys. Rev. Lett.},
  volume = {47},
  issue = {7},
  pages = {460},
  year = {1981},
  publisher = {American Physical Society},
  doi = {10.1103/PhysRevLett.47.460},
  url = {https://link.aps.org/doi/10.1103/PhysRevLett.47.460}
}

@article{Fabbrichesi:2022ovb,
  title={Constraining new physics in entangled two-qubit systems: top-quark, tau-lepton and photon pairs},
  author={Marco Fabbrichesi and Roberto Floreanini and Emidio Gabrielli},
  journal={The European Physical Journal C},
  volume={83},
  number={162},
  year={2023},
  publisher={Springer},
  url={https://link.springer.com/article/10.1140/epjc/s10052-023-11307-2}
}

@article{Barr:2021zcp,
  title={Testing Bell inequalities in Higgs boson decays},
  author={Alan J. Barr},
  journal={Physics Letters B},
  volume={825},
  pages={136866},
  year={2022},
  publisher={Elsevier},
  url={https://www.sciencedirect.com/science/article/pii/S0370269321008066}
}

@article{Barr:2022wyq,
  doi = {10.22331/q-2023-07-27-1070},
  url = {https://doi.org/10.22331/q-2023-07-27-1070},
  title = {Bell-type inequalities for systems of relativistic vector bosons},
  author = {Barr, Alan J. and Caban, Pawe{\l{}} and Rembieli{\'{n}}ski, Jakub},
  journal = {{Quantum}},
  issn = {2521-327X},
  publisher = {{Verein zur F{\"{o}}rderung des Open Access Publizierens in den Quantenwissenschaften}},
  volume = {7},
  pages = {1070},
  month = jul,
  year = {2023}
}

@article{Perotti:2018wxm,
  title = {Polarization observables in ${e}^{+}{e}^{\ensuremath{-}}$ annihilation to a baryon-antibaryon pair},
  author = {E. Perotti and G. F\"aldt and A. Kupsc and S. Leupold and J. J. Song},
  journal = {Phys. Rev. D},
  volume = {99},
  issue = {5},
  pages = {056008},
  year = {2019},
  publisher = {American Physical Society},
  doi = {10.1103/PhysRevD.99.056008},
  url = {https://link.aps.org/doi/10.1103/PhysRevD.99.056008}
}

@article{Ehataht:2023zzt,
  title = {Probing entanglement and testing Bell inequality violation with ${e}^{+}{e}^{\ensuremath{-}}\ensuremath{\rightarrow}{\ensuremath{\tau}}^{+}{\ensuremath{\tau}}^{\ensuremath{-}}$ at Belle II},
  author = {Ehat\"aht, K. and Fabbrichesi, M. and Marzola, L. and Veelken, C.},
  journal = {Phys. Rev. D},
  volume = {109},
  issue = {3},
  pages = {032005},
  numpages = {14},
  year = {2024},
  month = {Feb},
  publisher = {American Physical Society},
  doi = {10.1103/PhysRevD.109.032005},
  url = {https://link.aps.org/doi/10.1103/PhysRevD.109.032005}
}

@article{Wootters:1998prl,
  title = {Entanglement of Formation of an Arbitrary State of Two Qubits},
  author = {Wootters, William K.},
  journal = {Phys. Rev. Lett.},
  volume = {80},
  issue = {10},
  pages = {2245--2248},
  numpages = {0},
  year = {1998},
  month = {Mar},
  publisher = {American Physical Society},
  doi = {10.1103/PhysRevLett.80.2245},
  url = {https://link.aps.org/doi/10.1103/PhysRevLett.80.2245}
}

@article{Faldt:2017kgy,
title = {Hadronic structure functions in the e+e−→Λ¯Λ reaction},
journal = {Physics Letters B},
volume = {772},
pages = {16-20},
year = {2017},
issn = {0370-2693},
doi = {https://doi.org/10.1016/j.physletb.2017.06.011},
url = {https://www.sciencedirect.com/science/article/pii/S0370269317304719},
author = {F{\"a}ldt, G{\"o}ran and Kupsc, Andrzej}
}

@misc{Wu:2024mtj,
      title={Generalized quantum measurement in spin-correlated hyperon-antihyperon decays}, 
      author={Sihao Wu and Chen Qian and Yang-Guang Yang and Qun Wang},
      year={2024},
      eprint={2402.16574},
      archivePrefix={arXiv},
      primaryClass={hep-ph}
}

@article{Aguilar-Saavedra:2022wam,
  title = {Testing entanglement and Bell inequalities in $H\ensuremath{\rightarrow}ZZ$},
  author = {Aguilar-Saavedra, J. A. and Bernal, A. and Casas, J. A. and Moreno, J. M.},
  journal = {Phys. Rev. D},
  volume = {107},
  issue = {1},
  pages = {016012},
  numpages = {12},
  year = {2023},
  month = {Jan},
  publisher = {American Physical Society},
  doi = {10.1103/PhysRevD.107.016012},
  url = {https://link.aps.org/doi/10.1103/PhysRevD.107.016012}
}

@article{BESIII:2021ypr,
  title={Probing CP symmetry and weak phases with entangled double-strange baryons},
  author={Ablikim, Medina and others},
  collaboration = {BESIII Collaboration},
  doi = {10.1038/s41586-022-04624-1},
  journal={Nature},
  volume={606},
  number={7912},
  pages={64--69},
  year={2022},
  publisher={Nature Publishing Group UK London}
}

@article{BESIII:2017kqw,
  title = {Study of $J/\ensuremath{\psi}$ and $\ensuremath{\psi}(3686)$ decay to $\mathrm{\ensuremath{\Lambda}}\overline{\mathrm{\ensuremath{\Lambda}}}$ and ${\mathrm{\ensuremath{\Sigma}}}^{0}{\overline{\mathrm{\ensuremath{\Sigma}}}}^{0}$ final states},
  author = {Ablikim, Medina and  others},
  collaboration = {BESIII Collaboration},
  journal = {Phys. Rev. D},
  volume = {95},
  issue = {5},
  pages = {052003},
  numpages = {10},
  year = {2017},
  month = {Mar},
  publisher = {American Physical Society},
  doi = {10.1103/PhysRevD.95.052003},
  url = {https://link.aps.org/doi/10.1103/PhysRevD.95.052003}
}

@article{BES:2008hwe,
  title = {First measurements of $J/\ensuremath{\psi}$ decays into ${\ensuremath{\Sigma}}^{+}{\overline{\ensuremath{\Sigma}}}^{\ensuremath{-}}$ and ${\ensuremath{\Xi}}^{0}{\overline{\ensuremath{\Xi}}}^{0}$},
  author = {Ablikim, M. and others},
  collaboration = {BES Collaboration},
  journal = {Phys. Rev. D},
  volume = {78},
  issue = {9},
  pages = {092005},
  numpages = {7},
  year = {2008},
  month = {Nov},
  publisher = {American Physical Society},
  doi = {10.1103/PhysRevD.78.092005},
  url = {https://link.aps.org/doi/10.1103/PhysRevD.78.092005}
}

@article{BESIII:2020fqg,
  title = {${\mathrm{\ensuremath{\Sigma}}}^{+}$ and ${\overline{\mathrm{\ensuremath{\Sigma}}}}^{\ensuremath{-}}$ Polarization in the $J/\ensuremath{\psi}$ and $\ensuremath{\psi}(3686)$ Decays},
  author = {Ablikim, M. and others},
  collaboration = {BESIII Collaboration},
  journal = {Phys. Rev. Lett.},
  volume = {125},
  issue = {5},
  pages = {052004},
  numpages = {8},
  year = {2020},
  month = {Jul},
  publisher = {American Physical Society},
  doi = {10.1103/PhysRevLett.125.052004},
  url = {https://link.aps.org/doi/10.1103/PhysRevLett.125.052004}
}

@article{BESIII:2016nix,
title = {Study of $J/\psi$ and $\psi(3686)\rightarrow\Sigma(1385)^{0}\bar\Sigma(1385)^{0}$ and $\Xi^0\bar\Xi^{0}$},
journal = {Phys. Lett. B},
volume = {770},
pages = {217-225},
year = {2017},
issn = {0370-2693},
doi = {https://doi.org/10.1016/j.physletb.2017.04.048},
url = {https://www.sciencedirect.com/science/article/pii/S0370269317303222},
author = {Ablikim, Medina and others}
}

@article{BESIII:2023drj,
  title = {Tests of $CP$ symmetry in entangled ${\mathrm{\ensuremath{\Xi}}}^{0}\ensuremath{-}{\overline{\mathrm{\ensuremath{\Xi}}}}^{0}$ pairs},
  author = {Ablikim, Medina and others},
  collaboration = {BESIII Collaboration},
  journal = {Phys. Rev. D},
  volume = {108},
  issue = {3},
  pages = {L031106},
  numpages = {10},
  year = {2023},
  month = {Aug},
  publisher = {American Physical Society},
  doi = {10.1103/PhysRevD.108.L031106},
  url = {https://link.aps.org/doi/10.1103/PhysRevD.108.L031106}
}

@article{Bernal:2023jba,
  title = {Quantum tomography of helicity states for general scattering processes},
  author = {Bernal, Alexander},
  journal = {Phys. Rev. D},
  volume = {109},
  issue = {11},
  pages = {116007},
  numpages = {22},
  year = {2024},
  month = {Jun},
  publisher = {American Physical Society},
  doi = {10.1103/PhysRevD.109.116007},
  url = {https://link.aps.org/doi/10.1103/PhysRevD.109.116007}
}

@inproceedings{schrodinger:1935discussion,
  title={Discussion of probability relations between separated systems},
  author={Schr{\"o}dinger, Erwin},
  booktitle={Mathematical Proceedings of the Cambridge Philosophical Society},
  volume={31},
  number={4},
  pages={555--563},
  year={1935},
  organization={Cambridge University Press},
  doi = {10.1017/S0305004100013554}
}

@article{Wiseman:2007prl,
  title = {Steering, Entanglement, Nonlocality, and the Einstein-Podolsky-Rosen Paradox},
  author = {Wiseman, H. M. and Jones, S. J. and Doherty, A. C.},
  journal = {Phys. Rev. Lett.},
  volume = {98},
  issue = {14},
  pages = {140402},
  numpages = {4},
  year = {2007},
  month = {Apr},
  publisher = {American Physical Society},
  doi = {10.1103/PhysRevLett.98.140402},
  url = {https://link.aps.org/doi/10.1103/PhysRevLett.98.140402}
}

@article{Du:2021pra,
  title = {Relationship between first-order coherence and the maximum violation of the three-setting linear steering inequality for a two-qubit system},
  author = {Du, Ming-Ming and Tong, D. M.},
  journal = {Phys. Rev. A},
  volume = {103},
  issue = {3},
  pages = {032407},
  numpages = {6},
  year = {2021},
  month = {Mar},
  publisher = {American Physical Society},
  doi = {10.1103/PhysRevA.103.032407},
  url = {https://link.aps.org/doi/10.1103/PhysRevA.103.032407}
}

@article{Das:2019pra,
  title = {Detecting Einstein-Podolsky-Rosen steering through entanglement detection},
  author = {Das, Debarshi and Sasmal, Souradeep and Roy, Sovik},
  journal = {Phys. Rev. A},
  volume = {99},
  issue = {5},
  pages = {052109},
  numpages = {10},
  year = {2019},
  month = {May},
  publisher = {American Physical Society},
  doi = {10.1103/PhysRevA.99.052109},
  url = {https://link.aps.org/doi/10.1103/PhysRevA.99.052109}
}

@article{Zhang:2021pra,
  title = {Asymmetric steerability of quantum equilibrium and nonequilibrium steady states through entanglement detection},
  author = {Zhang, Kun and Wang, Jin},
  journal = {Phys. Rev. A},
  volume = {104},
  issue = {4},
  pages = {042404},
  numpages = {16},
  year = {2021},
  month = {Oct},
  publisher = {American Physical Society},
  doi = {10.1103/PhysRevA.104.042404},
  url = {https://link.aps.org/doi/10.1103/PhysRevA.104.042404}
}

@article{Ollivier:2001prl,
  title = {Quantum Discord: A Measure of the Quantumness of Correlations},
  author = {Ollivier, Harold and Zurek, Wojciech H.},
  journal = {Phys. Rev. Lett.},
  volume = {88},
  issue = {1},
  pages = {017901},
  numpages = {4},
  year = {2001},
  month = {Dec},
  publisher = {American Physical Society},
  doi = {10.1103/PhysRevLett.88.017901},
  url = {https://link.aps.org/doi/10.1103/PhysRevLett.88.017901}
}

@article{Jing:2016jpa,
doi = {10.1088/1751-8113/49/38/385302},
url = {https://dx.doi.org/10.1088/1751-8113/49/38/385302},
year = {2016},
month = {aug},
publisher = {IOP Publishing},
volume = {49},
number = {38},
pages = {385302},
author = {Naihuan Jing and Bing Yu},
title = {Quantum discord of X-states as optimization of a one variable function},
journal = {Journal of Physics A: Mathematical and Theoretical}
}

@article{Dakic:2010prl,
  title = {Necessary and Sufficient Condition for Nonzero Quantum Discord},
  author = {Daki{\'c}, Borivoje and Vedral, Vlatko and Brukner, {\v{C}}aslav},
  journal = {Phys. Rev. Lett.},
  volume = {105},
  issue = {19},
  pages = {190502},
  numpages = {4},
  year = {2010},
  month = {Nov},
  publisher = {American Physical Society},
  doi = {10.1103/PhysRevLett.105.190502},
  url = {https://link.aps.org/doi/10.1103/PhysRevLett.105.190502}
}

@article{Girolami:2011pra,
  title = {Quantum discord for general two-qubit states: Analytical progress},
  author = {Girolami, Davide and Adesso, Gerardo},
  journal = {Phys. Rev. A},
  volume = {83},
  issue = {5},
  pages = {052108},
  numpages = {8},
  year = {2011},
  month = {May},
  publisher = {American Physical Society},
  doi = {10.1103/PhysRevA.83.052108},
  url = {https://link.aps.org/doi/10.1103/PhysRevA.83.052108}
}

@article{Cavalcanti:2009pra,
  title = {Experimental criteria for steering and the Einstein-Podolsky-Rosen paradox},
  author = {Cavalcanti, E. G. and Jones, S. J. and Wiseman, H. M. and Reid, M. D.},
  journal = {Phys. Rev. A},
  volume = {80},
  issue = {3},
  pages = {032112},
  numpages = {16},
  year = {2009},
  month = {Sep},
  publisher = {American Physical Society},
  doi = {10.1103/PhysRevA.80.032112},
  url = {https://link.aps.org/doi/10.1103/PhysRevA.80.032112}
}

@article{Costa:2016pra,
  title = {Quantification of Einstein-Podolsky-Rosen steering for two-qubit states},
  author = {Costa, A. C. S. and Angelo, R. M.},
  journal = {Phys. Rev. A},
  volume = {93},
  issue = {2},
  pages = {020103},
  numpages = {5},
  year = {2016},
  month = {Feb},
  publisher = {American Physical Society},
  doi = {10.1103/PhysRevA.93.020103},
  url = {https://link.aps.org/doi/10.1103/PhysRevA.93.020103}
}

@article{Wu:2024asu,
  title = {Bell nonlocality and entanglement in ${e}^{+}{e}^{\ensuremath{-}}\ensuremath{\rightarrow}Y\overline{Y}$ at BESIII},
  author = {Wu, Sihao and Qian, Chen and Wang, Qun and Zhou, Xiao-Rong},
  journal = {Phys. Rev. D},
  volume = {110},
  issue = {5},
  pages = {054012},
  numpages = {10},
  year = {2024},
  month = {Sep},
  publisher = {American Physical Society},
  doi = {10.1103/PhysRevD.110.054012},
  url = {https://link.aps.org/doi/10.1103/PhysRevD.110.054012}
}

@article{Fabbrichesi:2024rec,
  title = {Bell inequality is violated in charmonium decays},
  author = {Fabbrichesi, M. and Floreanini, R. and Gabrielli, E. and Marzola, L.},
  journal = {Phys. Rev. D},
  volume = {110},
  issue = {5},
  pages = {053008},
  numpages = {21},
  year = {2024},
  month = {Sep},
  publisher = {American Physical Society},
  doi = {10.1103/PhysRevD.110.053008},
  url = {https://link.aps.org/doi/10.1103/PhysRevD.110.053008}
}

@article{Luo:2008pra,
  title = {Quantum discord for two-qubit systems},
  author = {Luo, Shunlong},
  journal = {Phys. Rev. A},
  volume = {77},
  issue = {4},
  pages = {042303},
  numpages = {6},
  year = {2008},
  month = {Apr},
  publisher = {American Physical Society},
  doi = {10.1103/PhysRevA.77.042303},
  url = {https://link.aps.org/doi/10.1103/PhysRevA.77.042303}
}

@article{BESIII:2024nif,
  title = {Strong and Weak $CP$ Tests in Sequential Decays of Polarized ${\mathrm{\ensuremath{\Sigma}}}^{0}$ Hyperons},
  author = {Ablikim, M. and others},
  collaboration = {BESIII Collaboration},
  journal = {Phys. Rev. Lett.},
  volume = {133},
  issue = {10},
  pages = {101902},
  numpages = {10},
  year = {2024},
  month = {Sep},
  publisher = {American Physical Society},
  doi = {10.1103/PhysRevLett.133.101902},
  url = {https://link.aps.org/doi/10.1103/PhysRevLett.133.101902}
}

@article{Adesso:2016jpa,
doi = {10.1088/1751-8113/49/47/473001},
url = {https://dx.doi.org/10.1088/1751-8113/49/47/473001},
year = {2016},
month = {nov},
publisher = {IOP Publishing},
volume = {49},
number = {47},
pages = {473001},
author = {Gerardo Adesso and Thomas R Bromley and Marco Cianciaruso},
title = {Measures and applications of quantum correlations},
journal = {Journal of Physics A: Mathematical and Theoretical}
}

@article{Uola:2020rmp,
  title = {Quantum steering},
  author = {Uola, Roope and Costa, Ana C. S. and Nguyen, H. Chau and G\"uhne, Otfried},
  journal = {Rev. Mod. Phys.},
  volume = {92},
  issue = {1},
  pages = {015001},
  numpages = {40},
  year = {2020},
  month = {Mar},
  publisher = {American Physical Society},
  doi = {10.1103/RevModPhys.92.015001},
  url = {https://link.aps.org/doi/10.1103/RevModPhys.92.015001}
}

@article{Nguyen:2019prl,
  title = {Geometry of Einstein-Podolsky-Rosen Correlations},
  author = {Nguyen, H. Chau and Nguyen, Huy-Viet and G\"uhne, Otfried},
  journal = {Phys. Rev. Lett.},
  volume = {122},
  issue = {24},
  pages = {240401},
  numpages = {6},
  year = {2019},
  month = {Jun},
  publisher = {American Physical Society},
  doi = {10.1103/PhysRevLett.122.240401},
  url = {https://link.aps.org/doi/10.1103/PhysRevLett.122.240401}
}

@article{Zhu:2018qip,
	title = {Analytical expression of quantum discord for rank-2 two-qubit states},
	volume = {17},
	issn = {1573-1332},
	url = {https://doi.org/10.1007/s11128-018-2007-6},
	doi = {10.1007/s11128-018-2007-6},
	pages = {234},
	number = {9},
	journal = {Quantum Information Processing},
	author = {Zhu, Xue-Na and Fei, Shao-Ming and Li-Jost, Xianqing},
	year = {2018}
}

@article{Fanchini:2010pra,
  title = {Non-Markovian dynamics of quantum discord},
  author = {Fanchini, F. F. and Werlang, T. and Brasil, C. A. and Arruda, L. G. E. and Caldeira, A. O.},
  journal = {Phys. Rev. A},
  volume = {81},
  issue = {5},
  pages = {052107},
  numpages = {6},
  year = {2010},
  month = {May},
  publisher = {American Physical Society},
  doi = {10.1103/PhysRevA.81.052107},
  url = {https://link.aps.org/doi/10.1103/PhysRevA.81.052107}
}

@article{Ferraro:2010pra,
  title = {Almost all quantum states have nonclassical correlations},
  author = {Ferraro, A. and Aolita, L. and Cavalcanti, D. and Cucchietti, F. M. and Ac\'{\i}n, A.},
  journal = {Phys. Rev. A},
  volume = {81},
  issue = {5},
  pages = {052318},
  numpages = {6},
  year = {2010},
  month = {May},
  publisher = {American Physical Society},
  doi = {10.1103/PhysRevA.81.052318},
  url = {https://link.aps.org/doi/10.1103/PhysRevA.81.052318}
}

@misc{Du:2024sly,
      title={Impact of parity violation on quantum entanglement and Bell nonlocality}, 
      author={Yong Du and Xiao-Gang He and Chia-Wei Liu and Jian-Ping Ma},
      year={2024},
      eprint={2409.15418},
      archivePrefix={arXiv},
      primaryClass={hep-ph},
      url={https://arxiv.org/abs/2409.15418}, 
}

@article{Czachor:1997pra,
  title = {Einstein-Podolsky-Rosen-Bohm experiment with relativistic massive particles},
  author = {Czachor, Marek},
  journal = {Phys. Rev. A},
  volume = {55},
  issue = {1},
  pages = {72--77},
  numpages = {0},
  year = {1997},
  month = {Jan},
  publisher = {American Physical Society},
  doi = {10.1103/PhysRevA.55.72},
  url = {https://link.aps.org/doi/10.1103/PhysRevA.55.72}
}

@article{Gingrich:2002prl,
  title = {Quantum Entanglement of Moving Bodies},
  author = {Gingrich, Robert M. and Adami, Christoph},
  journal = {Phys. Rev. Lett.},
  volume = {89},
  issue = {27},
  pages = {270402},
  numpages = {4},
  year = {2002},
  month = {Dec},
  publisher = {American Physical Society},
  doi = {10.1103/PhysRevLett.89.270402},
  url = {https://link.aps.org/doi/10.1103/PhysRevLett.89.270402}
}

@article{Peres:2004rmp,
  title = {Quantum information and relativity theory},
  author = {Peres, Asher and Terno, Daniel R.},
  journal = {Rev. Mod. Phys.},
  volume = {76},
  issue = {1},
  pages = {93--123},
  numpages = {0},
  year = {2004},
  month = {Jan},
  publisher = {American Physical Society},
  doi = {10.1103/RevModPhys.76.93},
  url = {https://link.aps.org/doi/10.1103/RevModPhys.76.93}
}

@article{Rembielinski:2019qbe,
  title = {Relativistic chiral qubits, their time evolution, and correlations},
  author = {Rembieli\ifmmode \acute{n}\else \'{n}\fi{}ski, Jakub and Caban, Pawe\l{}},
  journal = {Phys. Rev. A},
  volume = {99},
  issue = {2},
  pages = {022320},
  numpages = {6},
  year = {2019},
  month = {Feb},
  publisher = {American Physical Society},
  doi = {10.1103/PhysRevA.99.022320},
  url = {https://link.aps.org/doi/10.1103/PhysRevA.99.022320}
}

@article{Kurashvili:2022ybg,
doi = {10.1088/1751-8121/aca7a0},
url = {https://dx.doi.org/10.1088/1751-8121/aca7a0},
year = {2022},
month = {dec},
publisher = {IOP Publishing},
volume = {55},
number = {49},
pages = {495303},
author = {Podist Kurashvili and Levan Chotorlishvili},
title = {Quantum discord and entropic measures of two relativistic fermions},
journal = {Journal of Physics A: Mathematical and Theoretical}
}

@article{Barr:2024djo,
title = {Quantum entanglement and Bell inequality violation at colliders},
journal = {Progress in Particle and Nuclear Physics},
volume = {139},
pages = {104134},
year = {2024},
issn = {0146-6410},
doi = {https://doi.org/10.1016/j.ppnp.2024.104134},
url = {https://www.sciencedirect.com/science/article/pii/S0146641024000383},
author = {Alan J. Barr and Marco Fabbrichesi and Roberto Floreanini and Emidio Gabrielli and Luca Marzola}
}

@article{ATLAS:2023fsd,
  title={Observation of quantum entanglement with top quarks at the ATLAS detector},
  author={ATLAS collaboration and others},
  journal={Nature},
  volume={633},
  number={8030},
  pages={542},
  year={2024},
  publisher={Nature Publishing Group},
  doi = {10.1038/s41586-024-07824-z}
}

@article{Fabbrichesi:2023idl,
  title = {Bell inequality is violated in ${B}^{0}\ensuremath{\rightarrow}J/\ensuremath{\psi}{K}^{*}(892{)}^{0}$ decays},
  author = {Fabbrichesi, M. and Floreanini, R. and Gabrielli, E. and Marzola, L.},
  journal = {Phys. Rev. D},
  volume = {109},
  issue = {3},
  pages = {L031104},
  numpages = {5},
  year = {2024},
  month = {Feb},
  publisher = {American Physical Society},
  doi = {10.1103/PhysRevD.109.L031104},
  url = {https://link.aps.org/doi/10.1103/PhysRevD.109.L031104}
}

@article{Ming:2020nyc,
	title = {Quantification of quantumness in neutrino oscillations},
	volume = {80},
	issn = {1434-6052},
	url = {https://doi.org/10.1140/epjc/s10052-020-7840-y},
	doi = {10.1140/epjc/s10052-020-7840-y},
	pages = {275},
	number = {3},
	journal = {The European Physical Journal C},
	author = {Ming, Fei and Song, Xue-Ke and Ling, Jiajie and Ye, Liu and Wang, Dong},
	date = {2020-03-26}
}

@article{Bittencourt:2022tcl,
	title = {Complete complementarity relations for quantum correlations in neutrino oscillations},
	volume = {82},
	issn = {1434-6052},
	url = {https://doi.org/10.1140/epjc/s10052-022-10508-5},
	doi = {10.1140/epjc/s10052-022-10508-5},
  pages = {566},
	number = {6},
	journal = {The European Physical Journal C},
	author = {Bittencourt, V. A. S. V. and Blasone, M. and De Siena, S. and Matrella, C.},
	date = {2022-06-27}
}

@article{Shi:2011jpa,
doi = {10.1088/1751-8113/44/41/415304},
url = {https://dx.doi.org/10.1088/1751-8113/44/41/415304},
year = {2011},
month = {sep},
publisher = {IOP Publishing},
volume = {44},
number = {41},
pages = {415304},
author = {Mingjun Shi and Wei Yang and Fengjian Jiang and Jiangfeng Du},
title = {Quantum discord of two-qubit rank-2 states},
journal = {Journal of Physics A: Mathematical and Theoretical}
}

@article{Datta:2008prl,
  title = {Quantum Discord and the Power of One Qubit},
  author = {Datta, Animesh and Shaji, Anil and Caves, Carlton M.},
  journal = {Phys. Rev. Lett.},
  volume = {100},
  issue = {5},
  pages = {050502},
  numpages = {4},
  year = {2008},
  month = {Feb},
  publisher = {American Physical Society},
  doi = {10.1103/PhysRevLett.100.050502},
  url = {https://link.aps.org/doi/10.1103/PhysRevLett.100.050502}
}

@article{Lanyon:2008prl,
  title = {Experimental Quantum Computing without Entanglement},
  author = {Lanyon, B. P. and Barbieri, M. and Almeida, M. P. and White, A. G.},
  journal = {Phys. Rev. Lett.},
  volume = {101},
  issue = {20},
  pages = {200501},
  numpages = {4},
  year = {2008},
  month = {Nov},
  publisher = {American Physical Society},
  doi = {10.1103/PhysRevLett.101.200501},
  url = {https://link.aps.org/doi/10.1103/PhysRevLett.101.200501}
}

@article{Dakic:2012nphys,
	title = {Quantum discord as resource for remote state preparation},
	volume = {8},
	issn = {1745-2481},
	url = {https://doi.org/10.1038/nphys2377},
	doi = {10.1038/nphys2377},
	number = {9},
	journal = {Nature Physics},
	author = {Daki{\'c}, Borivoje and others},
	month = {Sep},
	year = {2012},
	pages = {666--670}
}

@article{BESIII:2023euh,
    author = "Ablikim, Medina and others",
    collaboration = "BESIII",
    title = "{Measurement of \ensuremath{\Lambda} transverse polarization in e$^{+}$e$^{−}$ collisions at $ \sqrt{s} $ = 3.68 \ensuremath{-} 3.71 GeV}",
    doi = "10.1007/JHEP10(2023)081",
    journal = "JHEP",
    volume = "10",
    pages = "081",
    year = "2023",
    note = "[Erratum: JHEP 12, 080 (2023)]"
}

@article{BESIII:2022lsz,
  title = {Observation of ${\mathrm{\ensuremath{\Xi}}}^{\ensuremath{-}}$ hyperon transverse polarization in $\ensuremath{\psi}(3686)\ensuremath{\rightarrow}{\mathrm{\ensuremath{\Xi}}}^{\ensuremath{-}}{\overline{\mathrm{\ensuremath{\Xi}}}}^{+}$},
  author = {Ablikim, M. and others},
  collaboration = {BESIII Collaboration},
  journal = {Phys. Rev. D},
  volume = {106},
  issue = {9},
  pages = {L091101},
  numpages = {9},
  year = {2022},
  month = {Nov},
  publisher = {American Physical Society},
  doi = {10.1103/PhysRevD.106.L091101},
  url = {https://link.aps.org/doi/10.1103/PhysRevD.106.L091101}
}

@article{Chen:2023oqs,
    author = "Chen, Cheng and Yan, Bing and Xie, Ju-Jun",
    title = "{Cross Sections and the Electromagnetic Form Factors within the Extended Vector Meson Dominance Model}",
    eprint = "2312.16753",
    archivePrefix = "arXiv",
    primaryClass = "hep-ph",
    doi = "10.1088/0256-307X/41/2/021302",
    journal = "Chin. Phys. Lett.",
    volume = "41",
    number = "2",
    pages = "021302",
    year = "2024"
}

@article{Wan:2021ncg,
    author = "Wan, Junyao and Yang, Yongliang and Lu, Zhun",
    title = "{The electromagnetic form factors of $\Lambda _c$ hyperon in the vector meson dominance model}",
    doi = "10.1140/epjp/s13360-021-01919-6",
    journal = "Eur. Phys. J. Plus",
    volume = "136",
    number = "9",
    pages = "949",
    year = "2021"
}

@article{Yang:2019mzq,
  title = {Electromagnetic form factors of $\mathrm{\ensuremath{\Lambda}}$ hyperon in the vector meson dominance model},
  author = {Yang, Yongliang and Chen, Dian-Yong and Lu, Zhun},
  journal = {Phys. Rev. D},
  volume = {100},
  issue = {7},
  pages = {073007},
  numpages = {9},
  year = {2019},
  month = {Oct},
  publisher = {American Physical Society},
  doi = {10.1103/PhysRevD.100.073007},
  url = {https://link.aps.org/doi/10.1103/PhysRevD.100.073007}
}

@article{Yan:2023nlb,
  title = {Understanding oscillating features of the timelike nucleon electromagnetic form factors within the extending vector meson dominance model},
  author = {Yan, Bing and Chen, Cheng and Li, Xia and Xie, Ju-Jun},
  journal = {Phys. Rev. D},
  volume = {109},
  issue = {3},
  pages = {036033},
  numpages = {8},
  year = {2024},
  month = {Feb},
  publisher = {American Physical Society},
  doi = {10.1103/PhysRevD.109.036033},
  url = {https://link.aps.org/doi/10.1103/PhysRevD.109.036033}
}

@article{Chen:2024luh,
doi = {10.1088/1674-1137/ad9259},
url = {https://dx.doi.org/10.1088/1674-1137/ad9259},
year = {2025},
month = {feb},
publisher = {Chinese Physical Society and the Institute of High Energy Physics of the Chinese Academy of Sciences and the Institute of Modern Physics of the Chinese Academy of Sciences and IOP Publishing Ltd
				},
volume = {49},
number = {2},
pages = {023102},
author = {Chen, Cheng and Yan, Bing and Xie, Ju-Jun},
title = {The electromagnetic form factors and spin polarization of {\boldsymbol\Lambda _{\boldsymbol c}^{\bf +}} in the process {{\boldsymbol e}^{\bf +}{\boldsymbol e}^{\bf -}\bf\to {\boldsymbol\Lambda}_{\boldsymbol c}^{\bf +} \bar{\boldsymbol\Lambda}_{\boldsymbol c}^{\bf -}}*},
journal = {Chinese Physics C}
}

@article{Haidenbauer:2020wyp,
  title = {Hyperon electromagnetic form factors in the timelike region},
  author = {Haidenbauer, Johann and Mei\ss{}ner, Ulf-G. and Dai, Ling-Yun},
  journal = {Phys. Rev. D},
  volume = {103},
  issue = {1},
  pages = {014028},
  numpages = {12},
  year = {2021},
  month = {Jan},
  publisher = {American Physical Society},
  doi = {10.1103/PhysRevD.103.014028},
  url = {https://link.aps.org/doi/10.1103/PhysRevD.103.014028}
}

@article{Wang:2024qbl,
    author = "Wang, Yadi",
    collaboration = "BESIII",
    title = "{Recent results of baryon EM form factors at BESIII}",
    doi = "10.1051/epjconf/202429203002",
    journal = "EPJ Web Conf.",
    volume = "292",
    pages = "03002",
    year = "2024"
}

@article{Achasov:2023gey,
    author = "Achasov, M. and others",
    title = "{STCF conceptual design report (Volume 1): Physics \& detector}",
    eprint = "2303.15790",
    archivePrefix = "arXiv",
    primaryClass = "hep-ex",
    doi = "10.1007/s11467-023-1333-z",
    journal = "Front. Phys. (Beijing)",
    volume = "19",
    number = "1",
    pages = "14701",
    year = "2024"
}

@article{Hensen:2015ccp,
    author = "Hensen, B. and others",
    title = "{Loophole-free Bell inequality violation using electron spins separated by 1.3 kilometres}",
    eprint = "1508.05949",
    archivePrefix = "arXiv",
    primaryClass = "quant-ph",
    doi = "10.1038/nature15759",
    journal = "Nature",
    volume = "526",
    pages = "682--686",
    year = "2015"
}

@article{Giustina:2015yza,
    author = "Giustina, Marissa and others",
    title = "{Significant-Loophole-Free Test of Bell\textquoteright{}s Theorem with Entangled Photons}",
    eprint = "1511.03190",
    archivePrefix = "arXiv",
    primaryClass = "quant-ph",
    doi = "10.1103/PhysRevLett.115.250401",
    journal = "Phys. Rev. Lett.",
    volume = "115",
    number = "25",
    pages = "250401",
    year = "2015"
}

@article{Pearle:1970prd,
  title = {Hidden-Variable Example Based upon Data Rejection},
  author = {Pearle, Philip M.},
  journal = {Phys. Rev. D},
  volume = {2},
  issue = {8},
  pages = {1418--1425},
  numpages = {0},
  year = {1970},
  month = {Oct},
  publisher = {American Physical Society},
  doi = {10.1103/PhysRevD.2.1418},
  url = {https://link.aps.org/doi/10.1103/PhysRevD.2.1418}
}

@misc{Bell:1995,
title = {Atomic-cascade photons and quantum-mechanical nonlocality},
author = {Bell, J S},
place = {France},
year = {1995},
month = {Dec},
}

@article{Dai:2024cpc,
doi = {10.1088/1674-1137/ad3dde},
url = {https://dx.doi.org/10.1088/1674-1137/ad3dde},
year = {2024},
month = {jul},
publisher = {Chinese Physical Society and the Institute of High Energy Physics of the Chinese Academy of Sciences and the Institute of Modern Physics of the Chinese Academy of Sciences and IOP Publishing Ltd
                        },
volume = {48},
number = {7},
pages = {073003},
author = {Dai, Jianping and Li, Hai-Bo and Miao, Han and Zhang, Jianyu},
title = {Prospects to study hyperon-nucleon interactions at BESIII*},
journal = {Chinese Physics C}
}

@article{BESIII:2023clq,
  title = {First Study of Reaction ${\mathrm{\ensuremath{\Xi}}}^{0}n\ensuremath{\rightarrow}{\mathrm{\ensuremath{\Xi}}}^{\ensuremath{-}}p$ Using ${\mathrm{\ensuremath{\Xi}}}^{0}$-Nucleus Scattering at an Electron-Positron Collider},
  author = {Ablikim, M. and others},
  collaboration = {BESIII Collaboration},
  journal = {Phys. Rev. Lett.},
  volume = {130},
  issue = {25},
  pages = {251902},
  numpages = {9},
  year = {2023},
  month = {Jun},
  publisher = {American Physical Society},
  doi = {10.1103/PhysRevLett.130.251902},
  url = {https://link.aps.org/doi/10.1103/PhysRevLett.130.251902}
}

@article{BESIII:2024geh,
  title = {First Study of Antihyperon-Nucleon Scattering $\overline{\mathrm{\ensuremath{\Lambda}}}p\ensuremath{\rightarrow}\overline{\mathrm{\ensuremath{\Lambda}}}p$ and Measurement of $\mathrm{\ensuremath{\Lambda}}p\ensuremath{\rightarrow}\mathrm{\ensuremath{\Lambda}}p$ Cross Section},
  author = {Ablikim, M. and others},
  collaboration = {BESIII Collaboration},
  journal = {Phys. Rev. Lett.},
  volume = {132},
  issue = {23},
  pages = {231902},
  numpages = {10},
  year = {2024},
  month = {Jun},
  publisher = {American Physical Society},
  doi = {10.1103/PhysRevLett.132.231902},
  url = {https://link.aps.org/doi/10.1103/PhysRevLett.132.231902}
}

@article{BESIII:2023trh,
    author = "Ablikim, M. and others",
    collaboration = "BESIII",
    title = "{First measurement of \ensuremath{\Lambda}N inelastic scattering with \ensuremath{\Lambda} from e+e\ensuremath{-}\textrightarrow{}J/\ensuremath{\psi}\textrightarrow{}\ensuremath{\Lambda}\ensuremath{\Lambda}\textasciimacron{}}",
    eprint = "2310.00720",
    archivePrefix = "arXiv",
    primaryClass = "nucl-ex",
    doi = "10.1103/PhysRevC.109.L052201",
    journal = "Phys. Rev. C",
    volume = "109",
    number = "5",
    pages = "L052201",
    year = "2024"
}

@article{Schlosshauer:2019ewh,
    author = "Schlosshauer, Maximilian",
    title = "{Quantum decoherence}",
    eprint = "1911.06282",
    archivePrefix = "arXiv",
    primaryClass = "quant-ph",
    doi = "10.1016/j.physrep.2019.10.001",
    journal = "Phys. Rept.",
    volume = "831",
    pages = "1--57",
    year = "2019"
}

@article{Caban:2014pra,
  title = {Relativistic Einstein-Podolsky-Rosen correlations and localization},
  author = {Caban, Pawe\l{} and Rembieli\ifmmode \acute{n}\else \'{n}\fi{}ski, Jakub and Rybka, Patrycja and Smoli\ifmmode \acute{n}\else \'{n}\fi{}ski, Kordian A. and Witas, Piotr},
  journal = {Phys. Rev. A},
  volume = {89},
  issue = {3},
  pages = {032107},
  numpages = {9},
  year = {2014},
  month = {Mar},
  publisher = {American Physical Society},
  doi = {10.1103/PhysRevA.89.032107},
  url = {https://link.aps.org/doi/10.1103/PhysRevA.89.032107}
}

@book{Breuer:2002pc,
  title={The theory of open quantum systems},
  author={Breuer, Heinz-Peter and Petruccione, Francesco},
  year={2002},
  publisher={OUP Oxford}
}

@article{Lindblad:1976generators,
  title={On the generators of quantum dynamical semigroups},
  author={Lindblad, Goran},
  journal={Communications in mathematical physics},
  volume={48},
  number={2},
  pages={119--130},
  year={1976},
  publisher={Springer}
}

@article{Gorini:1975nb,
    author = "Gorini, Vittorio and Kossakowski, Andrzej and Sudarshan, E. C. G.",
    title = "{Completely Positive Dynamical Semigroups of N Level Systems}",
    reportNumber = "CPT-244-TEXAS, ORO-3992-200",
    doi = "10.1063/1.522979",
    journal = "J. Math. Phys.",
    volume = "17",
    pages = "821",
    year = "1976"
}

@book{Nielsen2015,
address={Cambridge},
title={{Quantum Computation} and {Quantum Information}},
publisher={Cambridge University Press},
author={{M. A. Nielsen and I. L. Chuang}},
year={2015},
url={https://www.cambridge.org/highereducation/books/quantum-computation-and-quantum-information/01E10196D0A682A6AEFFEA52D53BE9AE},
}

@article{Jacobs:2006cp,
    title={A straightforward introduction to continuous quantum measurement},
  author={Jacobs, Kurt and Steck, Daniel A},
  journal={Contemporary Physics},
  volume={47},
  number={5},
  pages={279--303},
  year={2006},
  publisher={Taylor \& Francis}
}

@misc{Bechtle:2025ugc,
    author = "Bechtle, Philip and Breuning, Cedric and Dreiner, Herbi K. and Duhr, Claude",
    title = "{A critical appraisal of tests of locality and of entanglement versus non-entanglement at colliders}",
    eprint = "2507.15947",
    archivePrefix = "arXiv",
    primaryClass = "hep-ph",
    reportNumber = "BONN-TH-2025-23",
    month = "7",
    year = "2025"
}

@misc{Abel:2025skj,
    author = "Abel, Steven A. and Dreiner, Herbi K. and Sengupta, Rhitaja and Ubaldi, Lorenzo",
    title = "{Colliders are Testing neither Locality via Bell's Inequality nor Entanglement versus Non-Entanglement}",
    eprint = "2507.15949",
    archivePrefix = "arXiv",
    primaryClass = "hep-ph",
    reportNumber = "BONN-TH-2025-22, IPPP/25/48",
    month = "7",
    year = "2025"
}

@misc{Fabbrichesi:2025psr,
    author = "Fabbrichesi, M. and Floreanini, R. and Marzola, L.",
    title = "{Local vs. nonlocal entanglement in top-quark pairs at the LHC}",
    eprint = "2505.02902",
    archivePrefix = "arXiv",
    primaryClass = "hep-ph",
    month = "5",
    year = "2025"
}

@misc{Low:2025aqq,
    author = "Low, Matthew",
    title = "{Addressing Local Realism through Bell Tests at Colliders}",
    eprint = "2508.10979",
    archivePrefix = "arXiv",
    primaryClass = "hep-ph",
    month = "8",
    year = "2025"
}

@article{Henderson:2001wrr,
    author = "Henderson, L. and Vedral, V.",
    title = "{Classical, quantum and total correlations}",
    eprint = "quant-ph/0105028",
    archivePrefix = "arXiv",
    doi = "10.1088/0305-4470/34/35/315",
    journal = "J. Phys. A",
    volume = "34",
    number = "35",
    pages = "6899",
    year = "2001"
}

@article{BESIII:2025vsr,
   title={Test of local realism via entangled $$\Lambda \bar{\Lambda }$$ system},
   volume={16},
   author = {Ablikim, Medina and others},
   ISSN={2041-1723},
   url={http://dx.doi.org/10.1038/s41467-025-59498-4},
   DOI={10.1038/s41467-025-59498-4},
   number={1},
   journal={Nature Communications},
   publisher={Springer Science and Business Media LLC},
   year={2025},
   month=may
}

@article{Tang:2025oav,
    author = "Tang, Aihong",
    title = "{Probing spin and lifetime correlations in entangled hyperon-antihyperon pairs}",
    eprint = "2507.18507",
    archivePrefix = "arXiv",
    primaryClass = "nucl-ex",
    doi = "10.1016/j.physletb.2025.139820",
    journal = "Phys. Lett. B",
    volume = "868",
    pages = "139820",
    year = "2025"
}

@article{Cheng:2025cuv,
  title = {Bell Inequality Violation of Light Quarks in Dihadron Pair Production at Lepton Colliders},
  author = {Cheng, Kun and Yan, Bin},
  journal = {Phys. Rev. Lett.},
  volume = {135},
  issue = {1},
  pages = {011902},
  numpages = {7},
  year = {2025},
  month = {Jul},
  publisher = {American Physical Society},
  doi = {10.1103/gmqz-v4cl},
  url = {https://link.aps.org/doi/10.1103/gmqz-v4cl}
}

@article{Han:2024ugl,
    author = "Han, Tao and Low, Matthew and McGinnis, Navin and Su, Shufang",
    title = "{Measuring quantum discord at the LHC}",
    primaryClass = "hep-ph",
    reportNumber = "PITT-PACC-2316",
    doi = "10.1007/JHEP05(2025)081",
    journal = "JHEP",
    volume = "05",
    pages = "081",
    year = "2025"
}

@misc{Han:2025ewp,
    author = "Han, Tao and Low, Matthew and Su, Youle",
    title = "{Entanglement and Bell Nonlocality in $\tau^+ \tau^-$ at the BEPC}",
    eprint = "2501.04801",
    archivePrefix = "arXiv",
    primaryClass = "hep-ph",
    reportNumber = "PITT-PACC-2412",
    month = "1",
    year = "2025"
}

@misc{Pei:2025yvr,
    author = "Pei, Junle and Hao, Xiqing and Wang, Xiaochuan and Li, Tianjun",
    title = "{Observation of quantum entanglement in $\Lambda \bar{\Lambda}$ pair production via electron-positron annihilation}",
    eprint = "2505.09931",
    archivePrefix = "arXiv",
    primaryClass = "hep-ph",
    month = "5",
    year = "2025"
}

@article{Demina:2024dst,
  title = {Locality in collider tests of quantum mechanics with top quark pairs},
  author = {Demina, Regina and Landi, Gabriel},
  journal = {Phys. Rev. D},
  volume = {111},
  issue = {1},
  pages = {012013},
  numpages = {7},
  year = {2025},
  month = {Jan},
  publisher = {American Physical Society},
  doi = {10.1103/PhysRevD.111.012013},
  url = {https://link.aps.org/doi/10.1103/PhysRevD.111.012013}
}

@article{CMS:2024zkc,
    author = "Hayrapetyan, Aram and others",
    collaboration = "CMS",
    title = "{Measurements of polarization and spin correlation and observation of entanglement in top quark pairs using lepton+jets events from proton-proton collisions at s=13{\,}{\,}TeV}",
    eprint = "2409.11067",
    archivePrefix = "arXiv",
    primaryClass = "hep-ex",
    reportNumber = "CMS-TOP-23-007, CERN-EP-2024-231",
    doi = "10.1103/PhysRevD.110.112016",
    journal = "Phys. Rev. D",
    volume = "110",
    number = "11",
    pages = "112016",
    year = "2024"
}

@misc{Fabbrichesi:2025aqp,
    author = "Fabbrichesi, M. and Floreanini, R. and Marzola, L.",
    title = "{About testing Bell locality at colliders}",
    eprint = "2503.18535",
    archivePrefix = "arXiv",
    primaryClass = "quant-ph",
    month = "3",
    year = "2025"
}

@misc{Aoude:2025ovu,
    author = "Aoude, Rafael and Barr, Alan J. and Maltoni, Fabio and Satrioni, Leonardo",
    title = "{Decoherence effects in entangled fermion pairs at colliders}",
    eprint = "2504.07030",
    archivePrefix = "arXiv",
    primaryClass = "quant-ph",
    month = "4",
    year = "2025"
}

@misc{Gu:2025ijz,
    author = "Gu, Jiayin and Lin, Shi-Jia and Shao, Ding Yu and Wang, Lian-Tao and Yang, Si-Xiang",
    title = "{Decoherence in high energy collisions as renormalization group flow}",
    eprint = "2510.13951",
    archivePrefix = "arXiv",
    primaryClass = "hep-ph",
    month = "10",
    year = "2025"
}

\end{document}